\documentclass[acmtocl,acmnow]{acmtrans2m}

\usepackage{amsmath}
\usepackage{amssymb}
\usepackage{proof_tree}
\usepackage{times}
\usepackage{url}

\newtheorem{theorem}{Theorem}[section]

\newtheorem{corollary}[theorem]{Corollary}
\newtheorem{proposition}[theorem]{Proposition}
\newtheorem{lemma}[theorem]{Lemma}
\newdef{definition}[theorem]{Definition}
\newdef{remark}[theorem]{Remark}

\newdef{example}[theorem]{Example}

\newcommand{\comment}[1]{}

\newcommand{\card}[1]{\sharp{#1}}
\newcommand{\var}{\mathcal{V}}
\newcommand{\dom}{\operatorname{dom}}
\newcommand{\mgu}{\operatorname{mgu}}
\newcommand{\set}[1]{\lbrace{#1}\rbrace}
\newcommand{\subst}[2]{{}^{#2}\!/_{\!#1}}
\newcommand{\pair}[2]{\langle{#1},{#2}\rangle}
\newcommand{\st}{\mathit{St}}

\newcommand{\lleft}[1]{\mathsf{lhs}(#1)}
\newcommand{\lhs}{\mathsf{lhs}}

\newcommand{\rright}[1]{\mathsf{rhs}(#1)}
\newcommand{\rhs}{\mathsf{rhs}}
\newcommand{\ttrue}{\mathsf{\bf true}}
\newcommand{\HK}{K}
\newcommand{\simple}{partial}

\newcommand{\penc}[2]{\operatorname{enc}(#1,#2)}

\newcommand{\enc}{\operatorname{enc}}
\newcommand{\enca}{\operatorname{enca}}
\newcommand{\sign}{\operatorname{sign}}
\newcommand{\pub}[1]{#1}
\newcommand{\priv}{\operatorname{priv}}
\newcommand{\ar}{\operatorname{ar}}

\newcommand{\term}{\mathsf{Msg}}
\newcommand{\keys}{\mathsf{Key}}

\newcommand{\Time}{\mathsf{Time}}
\newcommand{\eqdef}{\stackrel{\mathsf{def}}{=}}
\newcommand{\rhoun}{\,\rho_1\,}

\newcommand{\lset}[1]{#1_s}

\newcommand{\hidden}[1]{\mathsf{hidden}({#1})}

\newcommand{\simpl}{\rightsquigarrow}   
\newcommand{\msimpl}{\rightsquigarrow}  
\newcommand{\gsimpl}{\leadsto}          

\newcommand{\sacyclic}{strictly acyclic}

\newcommand{\posp}{\operatorname{Pos_p}} 
\newcommand{\sko}{\prec}
\newcommand{\nko}{\preceq}

\newcommand{\logic}{{\cal L}}

\newcommand{\new}{\mathsf{generate}}
\newcommand{\send}{\mathsf{send}}
\newcommand{\recv}{\mathsf{receive}}

\newcommand{\dedcons}[1]{deducibility constraint}
\newcommand{\dedsys}[1]{deducibility constraint system}
\newcommand{\dedconss}[1]{deducibility constraints}
\newcommand{\dedsyss}[1]{deducibility constraint systems}
\newcommand{\Dedsyss}[1]{Deducibility constraint systems}

\newcommand{\avispa}{AVISPA}

\title{Deciding security properties for cryptographic protocols. Application to key cycles}

\author{
HUBERT COMON-LUNDH\\
ENS CACHAN \& Research Center for Information Security, AIST, Tokyo
\and
V\'ERONIQUE CORTIER\\
LORIA, CNRS \& Universit\'e Henri Poincar\'e \& INRIA Project CASSIS
\and 
EUGEN Z\u ALINESCU\\
MSR-INRIA Joint Centre, Orsay
}

\begin{abstract}
There is a large amount of work dedicated to the formal verification
of security protocols.
In this paper, we revisit and extend the NP-complete decision
procedure for a bounded number of sessions.
We use a, now standard, \dedcons{} formalism 
for modeling security protocols.
Our first contribution is to give a simple set of constraint simplification
rules, that allows to reduce any \dedsys{} to a set of \emph{solved forms}, representing all solutions (within the bound on sessions).

As a consequence, we prove that deciding the existence of key cycles is
NP-complete for a bounded number of sessions.
The problem of key-cycles has been put forward by recent works relating
computational and symbolic models.
The so-called \emph{soundness} of the symbolic model requires indeed
that no key cycle (e.g., $\enc(k,k)$) ever occurs in the execution of
the protocol. Otherwise, stronger security assumptions (such as KDM-security)
are required.

We show that our decision procedure can also be applied to prove again
the decidability
of authentication-like properties and the decidability of a significant 
fragment of protocols with timestamps.
\end{abstract}




\begin{document}

\begin{bottomstuff}
This work has been partially supported by the ACI-SI Satin and the ARA SSIA Formacrypt.
\end{bottomstuff}

\maketitle

\section{Introduction}
Security protocols are small programs that aim at securing communications over a public network, like
Internet.
Considering the increasing size of networks and their dependence on
cryptographic protocols, a high level of assurance is needed in the
correctness of such protocols.
The design of such protocols is difficult and error-prone; many attacks are discovered even
several years after the publication of a protocol.
Consequently, there has been 
a growing interest in applying formal methods for validating
cryptographic protocols and many results have been obtained.
The main advantage of this approach is its relative simplicity which makes it
amenable to automated analysis.
For example, the
secrecy preservation is co-NP-complete for a bounded number of sessions~\cite{AL00,RT01},
and decidable for an unbounded number of
sessions under some additional restrictions~\cite{rta03,durgin99undecidability,lowe98towards,ramanujam05jcs}.
Many tools  have also been developed
to automatically verify cryptographic protocols,
like~\cite{avispa2005,Blanchet_CSFW01_efficient_verifier,MS01,cremers08cav}.

\paragraph{Generalizing the constraint system approach} 
In this paper, we re-investigate and extend the NP-complete decision
procedure for a bounded number of sessions~\cite{RT01}. In this
setting (i.e. finite number of sessions), \dedsyss{} have
become the standard model for verifying security properties, with a
special focus on secrecy. Starting with Millen and Shmatikov's
paper~\cite{MS01} many results
(e.g.~\cite{CS03,BaudetCCS05,BCD-stacs2007}) have been obtained and
several tools (e.g.~\cite{CorinE02}) have been developed within this
framework.  Our first contribution is to provide a generic approach
derived from~\cite{CS03} to decide general security properties. We
show that any \dedsys{} can be transformed in (possibly
several) much simpler \dedsyss{} that are called \emph{solved
forms}, preserving \emph{all} solutions of the original system, and not
only its satisfiability. In other words, the \dedsys{} represents
in a symbolic way all the possible sequences of messages that are
produced, following the protocol rules, whatever are the intruder's actions.
This set of symbolic traces is infinite in general. Solved forms
are a simple (and finite) representation of such traces and we show that it is
suitable for the verification of many security properties. 
We also consider sorted terms, symmetric and asymmetric encryption, pairing and
signatures, but we do not consider algebraic properties like Abelian
groups or exclusive or.  In addition, we prove termination in
\emph{polynomial time} of the (non-deterministic) \dedcons{} simplification.
Compared to~\cite{RT01}, our procedure preserves all solutions. Hence,
we can represent for instance, all attacks on the secrecy and not
only decide if there exists one. Moreover,
presenting the decision procedure using a
small set of simplification rules yields more flexibility
for further extensions and modifications.

The main originality is that the method is applicable to any security
property that can be expressed as a formula on the protocol trace
and the agent memories.  For example, our decision procedure
(published in the LPAR'06 proceedings~\cite{CortierLPAR06}) has been
used in~\cite{CKKW-fsttcs2006} for proving that a new notion of
secrecy in presence of hashes is decidable (and co-NP-complete) for a
bounded number of sessions. It has also been used in \cite{cortier07fsttcs}
in the proof of modularity results for security of protocols.
To illustrate the large applicability of
our decision procedure, we show in this paper how it can be used for
proving co-NP-completeness of three kinds of security properties: the
existence of key cycles, authentication-like properties, and secrecy of
protocols with timestamps.

For authentication properties, we introduce a small logic that allows
to specify authentication and some similar security properties.  Using
our solved forms, we show that any property that can be
expressed within this logic can be decided.  The logic is smaller than
NPATRL~\cite{SyversonM96} or
$\mathcal{PS}$-LTL~\cite{corin-psltl,Corin-thesis}, but we believe
that decidability holds for a larger logic, closer to the two
above ones. However, the goal of this work is not to introduce a new logic,
but rather to highlight the proof method. Note also that the absence
of key cycles cannot be expressed in any of the three mentioned logics
because it is not only a trace property but also a property of the
message structure (see below).

For timestamps, we actually retrieve a significant fragment of
the decidable class identified by Bozga \textit{et al}~\cite{BEL04concur}.
We believe that our result can lead more easily to an implementation, since
we only need to adapt the procedure implemented in
\avispa~\cite{avispa2005}, while Bozga \textit{et al} have designed a
completely new decision procedure, which \textit{de facto} has not been implemented.

\paragraph{Application to key cycles} 
Our second main contribution is to use this approach to provide an
NP-complete decision procedure for detecting the generation of key
cycles during the execution of a protocol, in the presence of an
intruder, for a bounded number of sessions. To the best of our
knowledge, this problem has not been addressed before. The key cycle
problem is a problem that arises from the cryptographic community.
Indeed, two distinct approaches for the rigorous design and analysis
of cryptographic protocols have been pursued in the literature: the
so-called Dolev-Yao, symbolic, or formal approach on the one hand and
the cryptographic, computational, or concrete approach on the other
hand. In the symbolic approach, messages are modeled as formal terms
that the adversary can manipulate using a fixed set of operations. In
the cryptographic approach, messages are bit strings and the adversary
is an arbitrary probabilistic polynomial-time Turing machine. While
results in this model yield strong security guarantees, the proofs are
often quite involved and only rarely suitable for automation (see,
e.g.,
\cite{Goldwasser_Micali_JCSS84_probabilistic_encryption,BellareRogaway-CRYPTO-1993}).

Starting with the seminal work of Abadi and Rogaway
\cite{ARCryptology02}, recent results investigate the possibility of
bridging the gap between the two approaches.  The goal is to obtain
the best of both worlds: simple, automated security proofs that entail
strong security guarantees.  The approach usually consists in proving
that the Dolev-Yao abstraction of cryptographic primitives 
 is correct as soon as strong enough primitives are used in the
implementation.  For example, in the case of asymmetric encryption, it
has been shown~\cite{Micciancio_Warinschi_TCC04_soundness_of_formal_encryption}
that the perfect encryption assumption is a sound abstraction for
IND-CCA2, which corresponds to a well-established security level.  The
perfect encryption assumption intuitively states that encryption is a
black-box that can be opened only when one has the inverse
key. Otherwise, no information can be learned from a cipher-text about
the underlying plain-text.

However, it is not always sufficient to find the right cryptographic
hypotheses. Formal models may need to be amended in order to be
correct abstractions of the cryptographic models. 
A  widely used requirement is to control how keys can
encrypt other keys.  In a passive setting, soundness
results~\cite{ARCryptology02,MW04} require that no \emph{key cycles}
can be generated during the execution of a protocol. Key cycles are
messages like $\enc(k,k)$ or $\enc(k_1,k_2),\enc(k_2,k_1)$ where a key
encrypts itself or more generally when the encryption relation between
keys contains a cycle. Such key cycles have to be disallowed simply
because usual security definitions for encryption schemes do not
yield any guarantees otherwise.  In the active
setting, the typical hypotheses are even stronger.  For instance,
in~\cite{Backes_Pfitzmann_CSFW04_symmetric_encryption,cryptoeprint:2005:020}
the authors require that a key $k$ never encrypts a key generated
before $k$ or, more generally, that it is known in advance which key
encrypts which one. More precisely, the encryption relation has to be
compatible with the order in which keys are generated, or more
generally, it has to be compatible with an {a priori} given
\emph{ordering on keys}.

\paragraph{Related work on key cycles} 
Some authors circumvent the problem of key cycles by providing new
security definitions for encryption, \emph{Key Dependent Messages} security, or
KDM in short, that allow key
cycles~\cite{AdaoBanaHerzogScedrov-ESORICS05,backes07key}. However,
the standard security notions do not imply these new definitions, and
ad-hoc encryption schemes have to be constructed. 
Most of these constructions use the random oracle model, which
is provably non implementable. Though there was some recent
progress  \cite{hofheinz08towards} towards constructing a KDM-secure
encryption scheme in the standard model,
%
none of the usual, implemented encryption schemes has
been proved to satisfy KDM-security.

In a passive setting, Laud~\cite{Laud-NORDSEC02} proposed a
modification of the Dolev-Yao model such that the new model is a sound
abstraction even in the presence of key cycles. In his model the
intruder's power is strengthened by adding new deduction
rules. With the new rules, from a message containing a key cycle, the
intruder can infer all  keys  involved in the cycle as well as
 the messages encrypted by
these keys. Subsequently, Janvier~\cite{Janvier-these} proved that the
intruder deduction problem remains polynomial for the modified
deduction system. It was also suggested that this approach can be
extended to active intruders and incorporated in existing tools,
though, to the best of our knowledge, this has not been completed yet.  
Note that the
definition of key cycles used in~\cite{Janvier-these} is more
permissive than in~\cite{ARCryptology02} (which is unnecessarily
restrictive) and it corresponds to the approach of
Laud~\cite{Laud-NORDSEC02}.

\paragraph{Deciding key cycles} 
In this paper, we provide an NP-complete decision procedure for
detecting the generation of key cycles during the execution of a
protocol, in the presence of an active intruder, for a bounded number
of sessions.  Our procedure works for all the above mentioned
definitions of key cycles:  strict key cycles (\textit{\`a
la} Abadi, Rogaway),  non-strict (\textit{\`a la} Laud) key cycles,
key orderings (\textit{\`a la} Backes).  We therefore provide a
necessary component for automated tools used in proving strong,
cryptographic security properties, using existing soundness results.
Since our approach is an extension of the transformation rules derived
from the result of~\cite{RT01}, we believe that our algorithm can be
easily implemented since it can be adapted from the associated
procedure, already implemented in \avispa~\cite{avispa2005} for
deciding secrecy and authentication properties.

\paragraph{Outline of the paper} 
The messages and the intruder capabilities are modeled in
Section~\ref{sec:syntax}. In Section~\ref{sec:systems}, we define
\dedsyss{} and show how they can be used to express protocol
executions. In Section~\ref{sec:property}, we define security
properties and their satisfaction. In
Section~\ref{sec:approach}, we show that the satisfaction
of any (in)security property can be non-deterministically, polynomially
reduced to the satisfiability of the same problem, this time on
simpler constraint systems.  The simplification rules derived
from~\cite{CS03} are provided in Section~\ref{sec:rules}. They are
actually not sufficient to ensure termination in polynomial time. Thus
we introduce in Section~\ref{sec:poly} a refined decision procedure,
which is correct, complete, and terminating in polynomial time. 
We show in Section~\ref{sec:cycles} how this approach can be used to
obtain our main result of NP-completeness for the decision 
of the  key cycles generation.  In Section~\ref{sec:auth}, we introduce a small logic to
express authentication-like properties and we show how our technique
can be used to decide any formula of this logic.  In
Section~\ref{sec:timestamps}, we show how it can be used to derive
NP-completeness for protocols with timestamps. Some concluding remarks
about further work can be found in Section~\ref{sec:conclusion}.

\section{Messages and intruder capabilities}\label{sec:syntax}
\subsection{Syntax}

Cryptographic primitives are represented by function symbols. More specifically, we consider a
\emph{signature} $(\mathcal{S}, \mathcal{F})$ consisting
in  a set of \emph{sorts} $\mathcal{S} = \{s,s_1\ldots\}$
and a set of \emph{function symbols} $\mathcal{F} = \{\enc, \enca, \sign, \langle\,\rangle, \priv\}$.
Each function symbol is associated with an \emph{arity}: $\ar$ is a mapping
from $\mathcal{F}$ to ${\cal S}^* \times {\cal S}$, which we write 
$\ar(f) = s_1\times \cdots \times s_n \rightarrow s$.
The four first function symbols in $\mathcal{F}$ are binary: for each of
them  there are $s_1,s_2,s\in {\cal S}$ such that
$\ar(f)= s_1 \times s_2 \rightarrow s$. The  last symbol is unary:
there are $s,s'\in {\cal S}$ such that $\ar(f)= s\rightarrow s'$.

 The symbol $\langle\,\rangle$ represents the pairing function. The
terms $\enc(m,k)$ and $\enca(m,k)$  represent respectively the message $m$ encrypted with the symmetric
(resp. asymmetric) key $k$. The term $\sign(m,k)$ represents the message $m$ signed by the key $k$. The
term $\priv(a)$ represents the private key of the agent $a$. For simplicity,
we confuse the agents names with their public key. (Or conversely, we claim
that agents identities are defined by their public keys).

 $\mathcal{N} = \{a,b\ldots\}$ is a set of \emph{names} and $\mathcal{X}=\{x,y\ldots\}$ is
a set of \emph{variables}. Each name and each variable is associated with 
a sort. We assume that there are infinitely many names and infinitely
many variables of each sort.

The set of
\emph{terms of sort $s$} is defined inductively by
$$\begin{array}{lcl@{\hspace{0.4cm}}l}
  t & ::= & & \hspace{-0.4cm} \text{term of sort }s\\
  & | & x & \text{variable } x \text{ of sort } s\\
  & | & a & \text{name } a \text{ of sort } s\\
  & | & f(t_1,\ldots,t_n) & \text{application of symbol }f \in
  \mathcal{F} \text{ such that } \ar(f) = s_1\times \cdots \times s_n \rightarrow s\\ &&&
\text{ and each } t_i \text{ is a term of sort } s_i.
\end{array}$$

We assume a special sort $\term$ that
subsumes all the other sorts: any term is of sort $\term$.

Sorts are mostly left unspecified in this paper. They can be used in
applications to express that certain operators can be applied only
to some restricted terms.
For example, we use sorts explicitly to express that messages are encrypted by atomic keys (only in Section~\ref{sec:cycles}),
and to represent timestamps (only in Section~\ref{sec:timestamps}).

As usual, we write $\var(t)$ 
for the set of variables 
occurring in $t$. 
For a  set $T$ of terms, $\var(T)$ denotes the union of the variables occurring in the terms of $T$.
A term $t$ is \emph{ground} or \emph{closed} if and only if $\var(t)=\emptyset$.
A \emph{position} or an \emph{occurrence} in a term $t$ is a sequence of positive integers corresponding
to paths starting from the root in the tree-representation of $t$.
For a term $t$ and a position $p$ in this term, $t|_{p}$ denotes the  subterm of $t$ at
position $p$.
We write $\st(t)$ and $\st(T)$ for the set of subterms of a term $t$,
and of a set of terms $T$, respectively.  The \emph{size} of a term
$t$, denoted $|t|$, is defined inductively as usual: $|t|=1$ if $t$ is
a variable or a name and $t=1+\sum_{i=1}^n|t_i|$ if
$t=f(t_1,\dots,t_n)$ for $f\in\mathcal{F}$.  If~$T$ is a set of terms
then~$|T|$ denotes the sum of the sizes of its elements. The
cardinality of a set~$T$ is denoted by~$\card{T}$. By abuse of
notation, we sometimes denote by $T,u$ the set $T\cup\set{u}$.

Substitutions are written $\sigma = \{ \subst{x_1}{t_1} ,\ldots, \subst{x_n}{t_n} \}$ with $\dom(\sigma)
= \{ x_1,\ldots,x_n\}$. We only consider \emph{well-sorted} substitutions, 
for which $x_i$ and $t_i$ have the same sort. $\sigma$ is \emph{closed} if and only if every $t_i$ is
closed.
The application of a substitution $\sigma$ to a term $t$ is written $\sigma(t)$ or $t\sigma$.
A {most general unifier} of two terms $u$ and $v$ is denoted by $\mgu(u,v)$.


\subsection{Intruder capabilities}
The ability of the intruder is modeled by the deduction rules
displayed in Figure~\ref{fig:deduction} and corresponds to the usual
Dolev-Yao rules. 

Pairing, signing, symmetric and asymmetric encryption are the \emph{composition}
rules. The other rules are \emph{decomposition rules}.
Intuitively, these deduction rules say that an intruder can compose
messages by pairing, encrypting, and signing messages provided she has
the corresponding keys and conversely, she can decompose messages by
projecting or decrypting provided she holds the decryption keys. For
signatures, the intruder is also able to \emph{verify} whether a
signature $\sign(m,k)$ and a message $m$ match (provided she has the
verification key), but this does not give rise to any new message: 
 this capability needs not to be represented in the deduction system.
We also consider an optional rule
\[
\frac{S\vdash \sign(x,y)}{S\vdash
x}
\]
that expresses the ability to retrieve the whole message from
its signature. This property may or may not hold depending on the
signature scheme, and that is why this rule is optional. Note that
this rule is necessary for
obtaining soundness properties w.r.t.~cryptographic digital signatures. Our
results will hold in both cases,
whether or not this rule is considered in the deduction relation.

\begin{figure}[t]
\[\begin{array}{rcrc}
\text{Pairing} &
\prooftree
S\vdash x\quad S\vdash y
\justifies
 S\vdash \pair{x}{y}
\endprooftree
\quad\quad\quad
&\text{Symmetric encryption} &
 \prooftree S\vdash x\quad S\vdash y \justifies
S\vdash \enc(x,y)
\endprooftree
\\ \\
\text{Asymmetric encryption} &
 \prooftree S\vdash x\quad S\vdash y \justifies
S\vdash \enca(x,y)
\endprooftree
\quad\quad\quad &
\text{Signing} &
 \prooftree S\vdash x\quad S\vdash y \justifies
S\vdash \sign(x,y)
\endprooftree
\\ \\
\text{Symmetric decryption} &
 \prooftree S\vdash \enc(x,y)\quad S\vdash y
\justifies S\vdash x
\endprooftree
&
\text{First Projection} &
\prooftree
S\vdash \pair{x}{y}
\justifies
S\vdash x
\endprooftree
\\ \\
 \text{Asymmetric decryption} &
\prooftree
S\vdash \enca(x,\pub{y})\quad S\vdash \priv(y)
\justifies
S\vdash x
\endprooftree
&
\text{Second Projection} &
 \prooftree S\vdash \pair{x}{y} \justifies S\vdash y
\endprooftree
\\ \\
\text{Unsigning} \mbox{\it (optional)} &
 \prooftree S\vdash \sign(x,y) \justifies
S\vdash x \using 
\endprooftree
\quad\quad
&
 \text{Axiom} &
\prooftree
\justifies S,x\vdash x 
\endprooftree
\end{array}\]
\caption{Intruder deduction system.}
\label{fig:deduction}
\end{figure}

A \emph{proof tree} (sometimes simply called a proof) is a tree whose
labels are sequents $T\vdash u$ where $T$ is a finite set of terms
and $u$ is a term. A proof tree is inductively
defined as follows:
\begin{itemize}
\item if $u$ is a term and $u\in T$, then $T\vdash u$ is a proof tree
whose conclusion is $T\vdash u$, using the axiom;
\item if $\pi_1,\ldots,\pi_n$ are proof trees, whose respective conclusions
are $T\vdash u_1, \ldots, T\vdash u_n$ respectively and
$ \prooftree S\vdash t_1\quad\cdots\quad S\vdash t_n 
\justifies S\vdash t \endprooftree$
is a rule $R$ of the Figure~\ref{fig:deduction} such
that, for some (well-sorted) substitution $\sigma$, 
$t_1\sigma=u_1,\ldots,t_n\sigma=u_n$, then  $\prooftree 
\pi_1\quad \cdots\quad \pi_n  \justifies T\vdash t\sigma\endprooftree$ 
is a proof tree using $R$, whose conclusion is $T\vdash t\sigma$.
\end{itemize}
We will call \emph{subproof} a subtree of a proof tree. An \emph{strict subproof}
(resp. \emph{immediate subproof}) of $\pi$ is a subproof of $\pi$ distinct
from $\pi$ (resp. a maximal strict subproof of $\pi$).

A term $u$ is \emph{deducible} from a set of terms $T$, which we sometimes
write $T\vdash u$ by abuse of notation, if there exists a proof tree
whose conclusion is $T\vdash u$.

\begin{example}
The term $\pair{k_1}{k_2}$ is deducible from the set $S_1=\{\enc(k_1,k_2),k_2\}$,
as the following proof tree shows:
\[\prooftree
\prooftree
S_1\vdash \enc(k_1,k_2)\quad S_1\vdash k_2
\justifies
S_1\vdash k_1
\endprooftree
\quad
S_1\vdash k_2
\justifies
S_1\vdash \pair{k_1}{k_2}
\endprooftree
\]
\end{example}

\section{\Dedsyss{} and security properties}
Deducibility constraint systems
are quite common (see e.g.~\cite{MS01,CS03}) in modeling security protocols. 
 We recall
here their definition and show how they can be used to specify general security properties. Then we prove that
any \dedsys{} can be transformed into  simpler ones, called \emph{solved}. Such simplified constraints are then used to decide the security
properties.

\subsection{\Dedsyss{}}\label{sec:systems}

In the usual attacker's model, the intruder controls the network. In particular
she can schedule the messages. Once such a scheduling is fixed, she can
still replace the messages with fake ones, which are nevertheless accepted
by the honest participants. More precisely, some pieces of messages
cannot be analyzed by the participants, hence can be replaced by any
other piece, provided that the attacker can construct the overall message. 
This can be used to mount attacks. 

In the formal model, pieces that cannot be analyzed are replaced with variables.
Any substitution of these variables will be accepted, provided that the
attacker can deduce (using the deduction system of Figure~\ref{fig:deduction})
the corresponding instance. The main problem then is to decide whether
there is such a substitution, yielding a violation of the security property.

Let us give a detailed example recalling how possible execution traces are
formalized.

\begin{example}\label{ex:ns}
Consider the famous Needham-Schroeder asymmetric key
authentication protocol~\cite{NS78} designed for mutual authentication:
\[
\begin{array}{r@{\quad}l}
A\rightarrow B: & \enca(\pair{N_A}{A},\pub{B})\\
B\rightarrow A: & \enca(\pair{N_A}{N_B},\pub{A})\\
A\rightarrow B: & \enca(N_B,\pub{B})
\end{array}
\]
The agent $A$ sends to $B$ his name and a fresh nonce (a randomly generated value) encrypted with the public
key of $B$. The agent $B$ answers by copying $A$'s nonce and adds a fresh nonce $N_B$, encrypted by $A$'s
public key. The agent $A$ acknowledges by forwarding $B$'s nonce encrypted by $B$'s public key. 

Formally, this protocol can be described using two roles $A$ and $B$.
The role $A$ has two parameters: $a,b$ (initiator and responder), and
is (informally) specified as follows:
\[
\begin{array}{rl}
A(a,b)\: :\;\; & \new(n_a)\\
A1. & \send(\enca(\pair{n_a}{a},\pub{b}))\\
A2. & \recv(\enca{\pair{n_a}{y}},\pub{a}) \rightarrow \send(\enca(y,\pub{b}))
\end{array}
\]
where $y$ is a variable: $a$ cannot check that this piece of the message is
a nonce generated by $b$. Hence it can be replaced by any term (or any term
of a given sort, depending on what we want to model). 

Similarly, the role of $B$ takes the two parameters $b,a$, and is specified as:
\[
\begin{array}{rl}
B(b,a)\: :\;\; & \new(n_b)\\
B1.& \recv(\enca(\pair{x}{a},\pub{b})) \rightarrow \send(\enca(\pair{x}{n_b},\pub{a}))\\
B2. & \recv(\enca(n_b,\pub{b}))
\end{array}
\]


Without loss of generality, we may assume that $\send$ actions are
performed as soon as the corresponding $\recv$ action is completed:
this is the best scheduling strategy for the attacker, who will get more
information for further computing fake messages. For this reason, we only need
to consider the possible scheduling of $\recv$ events.

Let $a,b$ be honest participants and $i$ be a corrupted one.
Consider one session $A(a,i)$ and one session $B(b,a)$.
There are three message deliveries to schedule: $A2,B1,B2$ and $B2$ has to
occur after $B1$. Assume the chosen scheduling is $B1,A2,B2$.
In this scenario, the possible sequences of message  delivery are
instances of $\enca(\pair{x}{a},\pub{b})$, $\enca(\pair{n_a}{y},\pub{a})$,
$\enca(n_b,\pub{b})$.  The variables $x,y$ can be replaced by any term, provided that
the attacker can build the corresponding instances from her knowledge at
the appropriate control point.

The initial intruder knowledge can be set to
$T_0 = \{a,b,i,\priv(i)\}$, including the private
key of the corrupted agent.

For the first message delivery, the attacker has to be able to build
the first message instance from this initial knowledge and the message
sent at step $A1$:
\begin{eqnarray}
T_1\eqdef T_0\cup\{\enca(\pair{n_a}{a},\pub{i})\} & \Vdash &
\enca(\pair{x}{a},\pub{b})\label{eq1}
\end{eqnarray}

This notation will be formally defined later on. Informally, this is
a formula, which is satisfied by a substitution $\sigma$ on $x$ if
$\enca(\pair{x}{a},\pub{b})\sigma$ is deducible from $T_1$, expressing
the ability of the intruder to construct $\enca(\pair{x}{a},\pub{b})\sigma$.

Then, the agent $b$ replies sending the corresponding instance 
$\enca(\pair{x}{n_b},\pub{a})$, which increases the attacker's knowledge,
hence enabling its use for building the next message; we get the second
\dedcons{}:
\begin{eqnarray}
T_2\eqdef T_1\cup\{\enca(\pair{x}{n_b},\pub{a})\} & \Vdash &
\enca(\pair{n_a}{y},\pub{a})\label{eq2}
\end{eqnarray}
Similarly, we construct a third \dedcons{} for the last message delivery:
\begin{eqnarray}
T_3\eqdef T_2\cup\{\enca(y,\pub{i})\} & \Vdash &
\enca(n_b,\pub{b})
\end{eqnarray}

\end{example}

\begin{definition}\label{def:constraint_sys}
A \emph{\dedsys{}} $C$ is a finite set of expressions
$T\Vdash u$,\linebreak called \emph{\dedconss{}}, where $T$ is a non empty set of terms, called the \emph{left-hand
side} of the \dedcons{} and
$u$ is a term, called the \emph{right-hand side} of the \dedcons{}, such that:
\begin{enumerate}
\item\label{def-item:cs1} the left-hand sides of all \dedconss{} are totally ordered by inclusion;
\item\label{def-item:cs2} if $x\in\var(T)$ for some $(T\Vdash u)\in C$ then \[T_x\eqdef\min\set{T'\mid (T'\Vdash u') \in C,
x\in\var(u')}\] exists and $T_x\subsetneq T$.
\end{enumerate}
\end{definition}

Informally, the first condition states that the intruder knowledge is
always increasing. The second condition expresses that variables abstract
pieces of \emph{received} messages: they have to occur first on the right
side of a constraint $T\Vdash u$, before occurring in some left side.  
Note that, due to point $(\ref{def-item:cs1})$, $T_x$ exists if and only if the set
$\{T'\mid (T'\Vdash u') \in C, x\in\var(u')\}$ is not empty. 
The linear ordering on left hand sides also implies the uniqueness of the
minimum. Hence $(\ref{def-item:cs2})$ can be restated equivalently as:
\[ (\ref{def-item:cs2}) \;\; \forall x\in \var(C), \:\exists\, (T\Vdash u)\in C, \;
x\in \var(u)\setminus\var(T) \]
In what follows, we may use this formulation instead.



The \emph{left-hand side} of a \dedsys{} $C$, denoted by $\lleft{C}$,
is the maximal left-hand side of the \dedconss{} of $C$.
The \emph{right-hand side} of a \dedsys{}
$C$, denoted by $\rright{C}$, is the set of right-hand sides of its
\dedconss{}. $\var(C)$ denotes the set of variables occurring in
$C$. $\bot$ denotes the unsatisfiable system. The \emph{size} of a
constraint system is defined as 
$|C|\eqdef |\lleft{C}\cup\rright{C}|$.

A \dedsys{} $C$ is also written as a conjunction of \dedconss{} $$C=\bigwedge_{1\le i\le
n}(T_i\Vdash u_i)$$ with $T_i\subseteq T_{i+1}$, for all $i$ with $1\le i\le n-1$. The second condition in

Definition~\ref{def:constraint_sys} then implies that if $x\in\var(T_i)$ then $\exists j<i$ such that $T_j=T_x$
and $T_j\subsetneq T_i$.


\begin{definition}
A \emph{solution} $\sigma$ of a \dedsys{} $C$ is a (well-sorted) ground substitution whose domain is $\var(C)$ and such that, for every $T\Vdash u\in C$, $T\sigma \vdash u\sigma$.
\end{definition}

\begin{example}
Coming back to Example~\ref{ex:ns}, the substitution 
 $\sigma_1 = \{\subst{x}{n_a},\subst{y}{n_b}\}$ is a solution of the \dedsys{} since
\[
\begin{array}{rcl}
 T_0\cup\{\enca(\pair{n_a}{a},\pub{i})\} & \vdash &
\enca(\pair{x}{a},\pub{b})\sigma_1\\
T_1\sigma_1\cup\{\enca(\pair{x}{n_b},\pub{a})\sigma_1\} & \vdash &
\enca(\pair{n_a}{y},\pub{a})\sigma_1\\
T_2\sigma_1\cup\{\enca(y,\pub{i})\sigma_1\} & \vdash &
\enca(n_b,\pub{b})
\end{array}
\]
\end{example}

\subsection{Security properties}\label{sec:property}

Deducibility constraint systems
represent in a symbolic and compact way a
possibly infinite set of traces (behaviors), which depend on the 
attacker's actions.
Security properties are formulas, that are interpreted over these 
traces. 


\begin{definition}
Given a set of predicate symbols together with their interpretation over
the set of ground terms,
a \emph{(in)security property} is a first-order formula $\phi$ built on these predicate
symbols. 
A \emph{solution} of $\phi$ is a ground substitution $\sigma$
of $\var(\phi)$ such that $\phi\sigma$ is true in the given interpretation.
(We also write $\sigma \models \phi$).

If $C$ is a \dedsys{} and $\phi$ is a (in)security property,
possibly sharing free variables with $C$,
a closed substitution $\sigma$ from $\var(\phi)\cup \var(C)$ is an \emph{attack
for $\phi$ and  $C$}, if is a solution of both $C$ and $\phi$. 

\end{definition}


\begin{example}\label{ex:classic}
If the security property is simply $\ttrue$ (which is always satisfied)
and the only sort is $\term$ then we find the usual
\dedsys{} satisfaction problem, whose satisfiability
 is known to be NP-complete~\cite{RT03TCS}.
\end{example}

\begin{example}\label{ex:secrecy}
Secrecy can be easily expressed by requiring that the secret data is not
deducible from the messages sent on the network.
We consider again the \dedsys{} $C_1$ defined in
Example~\ref{ex:ns}.
The (in)security property then expresses that $n_b$ is deducible:
$\phi$ is the \dedcons{} $T_3\Vdash n_b$. 
Note that we may view a constraint (system) as a first order formula.


Then the substitution $\sigma_1 = \{\subst{x}{n_a},\subst{y}{n_b}\}$
is an attack for $\phi$ and $C_1$ 
and corresponds to the attack found by
G.~Lowe~\cite{lowe96breaking}.
Note
that such a deduction-based property can be directly included in
the constraint system by adding a \dedcons{} $T_3\Vdash n_b$.
\end{example}

\begin{example}
\label{ex:auth}
Let us show here an example of authentication property.
Two agents $A$ and $B$ authenticate on some message $m$ if whenever
$B$ finishes a session \emph{believing} he has talked to $A$ then $A$
has indeed finished a session with $B$ and they share the same value
for $m$. Note that the agents $A$ and $B$ have in general a different
view of the message $m$, depending e.g. on which nonces they have generated
themselves and on which nonces they have received.
If $m_A$ denotes the view of $m$ from $A$ and $m_B$ the view of $m$
from $B$, then the insecurity property states that there is a trace in which
these two messages are distinct. 

Back to Example~\ref{ex:ns}, consider another scenario with two instances
of the role $A$: $A(a,i)$ and $A(a,b)$ and one instance of the role $B$:
$B(b,a)$. The attacker schedules the communications as in Example~\ref{ex:ns}: in particular the expected message delivery in $A(a,b)$ is
not scheduled (the message is not delivered). Then the \dedsys{}
$C'_1$ is identical to $C_1$, except that $T_0$ is replaced with
%
$T_0'=T_0\cup\{\enca(\pair{n_a'}{a},\pub{b})\}$. 
The nonce~$x$ received by $b$ should correspond to the nonce $n_a'$
sent by $a$ for $b$; we consider $m_A = n_a'$, $m_B = x$. 

The failure of authentication can be stated as the simple formula $x \neq n'_a$.
The substitution $\sigma_1$ defined in Example~\ref{ex:secrecy} is then
an attack,
since $b$ accepts the nonce $n_a$ instead of $n_a'$: $x\sigma_1\neq n'_a$. 
\end{example}

In Sections~\ref{sec:cycles},~\ref{sec:auth},~\ref{sec:timestamps} we provide with other examples 
corresponding to time constraints, more general authentication-like
properties, or to express that no key
cycles are allowed.

\section{Simplifying \dedsyss{}}\label{sec:approach}
Using simplification rules, solving \dedsyss{} can be reduced to solving simpler
constraint systems that we call solved.
One nice property of the transformation is that it works for any
security property.

\begin{definition}
A \dedsys{} is \emph{solved} if it is
 $\bot$ or each of its constraints are of

the form 
$T\Vdash x$, where $x$ is a variable.
\end{definition}
This definition corresponds to the notion of solved form in~\cite{CS03}.
Note that the empty \dedsys{} is solved.

Solved \dedsyss{} with the single sort $\term$ are
particularly simple in the case of the $\ttrue$ predicate since they
always have a solution, as noticed in~\cite{MS01}. Indeed, let $T_1$
be the smallest (w.r.t.~inclusion) left hand side of all constraints
of a \dedsys{}. From Definition~\ref{def:constraint_sys}, 
 $T_1$ is non empty and has no variables. Let $t\in
T_1$. Then the substitution $\theta$ defined by $x\theta=t$ for every
variable~$x$ is a solution since $T\vdash x\theta=t$ for any
constraint $T\Vdash x$ in the solved system.



\subsection{Simplification rules}\label{sec:rules}

The \emph{simplification rules} we consider are defined in
Figure~\ref{fig:rules}. For instance, the rule $R_1$  removes a redundant
constraint, i.e., when it is a logical consequence of smaller constraints.
The rule $R_3$ guesses some identity (confusion) between two sent sub-messages.

All the rules are in fact indexed by a substitution:
when there is no index then the identity substitution is implicitly
assumed. We write $C\simpl^n_{\sigma} C'$ if there are
$C_1,\dots, C_n$ with $n\ge 1$, $C'=C_n$,
$C\simpl_{\sigma_1} C_1\simpl_{\sigma_2} \dots
\simpl_{\sigma_n}C_n$, and
$\sigma=\sigma_1\sigma_2\dots\sigma_n$. We write
$C\simpl^*_{\sigma} C'$ if $C\simpl^n_{\sigma}
C'$ for some $n\geq 1$, or if $C'=C$ and $\sigma$ is the identity
substitution.

\begin{figure}[t]
\[\begin{array}{l@{\quad}r@{\hspace{0.2cm}}l@{\hspace{0.3cm}}l}
R_1 & C\,\wedge\,T\Vdash u &
\simpl\hspace{0.2cm}C
& \mbox{if } T\cup\set{x\mid (T'\Vdash x)\in C, T'\subsetneq T}\!\vdash\!u\\[0.1cm]
R_2 & C\,\wedge\,T\Vdash u & \simpl_\sigma
C\sigma\,\wedge\,T\sigma\Vdash u\sigma &
\mbox{if } \sigma=\mgu(t,u),\, t\in \st(T),\\
\vspace{-0.5mm}&&&\hspace{0.5cm} t\neq u,\ t, u \mbox{ not
variables}
\\[0.1cm]
R_3 & C\,\wedge\,T\Vdash u & \simpl_\sigma
C\sigma\,\wedge\,T\sigma\Vdash u\sigma &
 \mbox{if } \sigma=\mgu(t_1,t_2),\, t_1,t_2\in \st(T),\\
\vspace{-0.5mm}&&& \hspace{0.8cm}t_1\neq t_2,\ t_1,t_2 \mbox{ not variables}\\[0.1cm]
R'_3 & C\,\wedge\,T\Vdash u & \simpl_\sigma
C\sigma\,\wedge\,T\sigma\Vdash u\sigma &
 \mbox{if } \sigma=\mgu(t_2,t_3),\, \enca(t_1,t_2),\priv(t_3)\in \st(T),\\
\vspace{-0.5mm}&&& \hspace{0.8cm}t_2\neq t_3,\ t_2\mbox{ or }t_3 \mbox{ (or both) is a variable}\\[0.1cm]
R_4 & C\,\wedge\,T\Vdash u & \simpl\hspace{0.2cm}\bot &
\mbox{if } \var(T,u)=\emptyset \mbox{ and }
T\not\vdash u\\[0.1cm]
R_f & C\,\wedge\,T\Vdash f(u,v) &
\simpl\hspace{0.2cm}C\,\wedge\,T\Vdash u\,\wedge\,T\Vdash
v & \mbox{for } f\in\{\,\langle\,\rangle,\enc,\enca,\sign\}\\

\comment{ R_{\pair{}{}} & C\,\wedge\,T\Vdash \pair{u}{v} &
\simpl\hspace{0.2cm}C\,\wedge\,T\Vdash u\,\wedge\,T\Vdash
v
\\
R_{\enc} & C\,\wedge\,T\Vdash \penc{u}{v} & \simpl\hspace{0.2cm}C\,\wedge\,T\Vdash
u\,\wedge\,T\Vdash v
\\
R_{\enca} & C\,\wedge\,T\Vdash \enca(u,v) & \simpl\hspace{0.2cm}C\,\wedge\,T\Vdash
u\,\wedge\,T\Vdash v
\\
R_{\sign} & C\,\wedge\,T\Vdash \sign(u,v) & \simpl\hspace{0.2cm}C\,\wedge\,T\Vdash
u\,\wedge\,T\Vdash v
\\
}
\end{array}\]
\caption{Simplification rules.}
\label{fig:rules}
\end{figure}



\begin{example}
Let us consider the following \dedsys{} $C$:
\[\left\{\begin{array}{rcl}
T_1 & \Vdash&  \langle\,\enca(x,\pub{a}),\, \enca(y,\pub{a})\,\rangle \\
T_2 & \Vdash& k_1
  \end{array}\right.
\]
where $T_1 = \{a,\langle\enca(k_1,\pub{a}), \enca(k_2,\pub{a})\rangle\}$ and $T_2 =
T_1 \cup \{\enc(y,x)\}$.
The \dedsys{} $C$ can be simplified into
a solved form  using (for example) the following sequence
of simplification rules.\\
\[
C\stackrel{R_{\langle\rangle}}{\simpl}
\left\{\begin{array}{@{}r@{\:}c@{\:}l}
T_1& \Vdash& \enca(x,\pub{a})\\
T_1& \Vdash& \enca(y,\pub{a})\\
T_2 & \Vdash& k_1
\end{array}\right.
\stackrel{R_{\enca}}{\simpl}
\left\{\begin{array}{@{}r@{\:}c@{\:}l}
T_1& \Vdash& x\\
T_1& \Vdash& a\\
T_1& \Vdash& \enca(y,\pub{a})\\
T_2 & \Vdash& k_1
\end{array}\right.
\stackrel{R_1}{\simpl}
\left\{\begin{array}{@{}r@{\:}c@{\:}l}
T_1& \Vdash& x\\
T_1& \Vdash& \enca(y,\pub{a})\\
T_2 & \Vdash& k_1
\end{array}\right.
\]
since
$T_1 \vdash a$.
Let $\sigma = \mgu\big(\enca(k_1,\pub{a}),\; \enca(y,\pub{a})\big) = \{\subst{y}{k_1}\}$. We~have
\[\left\{\begin{array}{rcl}
T_1& \Vdash& x\\
T_1& \Vdash& \enca(y,\pub{a})\\
T_2 & \Vdash& k_1
\end{array}\right.
\stackrel{R_2}{\simpl}_{\sigma}
\left\{\begin{array}{rcl}
T_1& \Vdash& x\\
T_1& \Vdash& \enca(k_1,\pub{a})\\
T_2\sigma & \Vdash & k_1
\end{array}\right.
\stackrel{R_1}{\simpl}
\left\{\begin{array}{rcl}
T_1& \Vdash& x\\
T_2\sigma & \Vdash & k_1
\end{array}\right.
\stackrel{R_1}{\simpl}
\begin{array}{rcl}
T_1& \Vdash& x
\end{array}
\]
since $T_1\vdash \enca(k_1,\pub{a})$ and
$T_2\sigma \cup \{x\}\vdash k_1$.
Intuitively, it means that any substitution of the form $\{\subst{x}{m}, \subst{y}{k_1}\}$
such that $m$ is deducible from $T_1$ is solution of $C$.
\end{example}

The simplification rules are correct and complete: a
\dedsys{} $C$ has a solution, which is also a solution of a (in)security property $\phi$, if
and only if there exists a \dedsys{} $C'$ in solved form such that
$C\simpl^*_{\sigma} C'$ and there is a solution of both $C'$ and
$\phi\sigma$.
Note that several simplification rules can possibly be applied to a
given \dedsys{}.
\begin{theorem}\label{theo:general}
Let $C$ be a \dedsys{}, $\theta$ a substitution, and $\phi$ a (in)security property.
\begin{enumerate}
\item(Correctness) If $C\simpl^*_{\sigma} C'$ for some
  \dedsys{} $C'$
  and some substitution $\sigma$, and  if $\theta$ is an attack for 
$\phi\sigma$ and~$C'$, 
then $\sigma\theta$ is an attack for $\phi$ and~$C$.
\item(Completeness) If $\theta$ is an attack for $C$  and $\phi$, 
then there exist a \dedsys{} $C'$ in solved form
  and  substitutions $\sigma,\theta'$ such that $\theta
  =\sigma\theta'$, $C\simpl^*_{\sigma} C'$, and $\theta'$ is 
an attack for $C'$ and $\phi\sigma$. 
\item(Termination)  There is no infinite derivation sequence
$C \simpl_{\sigma_1}\! C_1\! \simpl_{\sigma_2}\! \cdots\! \simpl_{\sigma_n}\! C_n \cdots$.
\end{enumerate}
\end{theorem}



Theorem~\ref{theo:general} is
proved in Sections~\ref{app:corect},~\ref{app:complet},~and~\ref{section:termination}.

Getting a polynomial bound on the length of simplification sequences
requires however an additional memorization technique.
This is explained in Section~\ref{section:complexity}.

%


\subsection{Correctness}\label{app:corect}

We first give two simple lemmas.
\begin{lemma}\label{lemma_deduc-var}
If $T\vdash u$ then $\var(u)\subseteq\var(T)$.
\end{lemma}
\begin{proof}
The statement follows by induction on the depth of a proof of $T\vdash u$,
observing that no deduction rule introduces new variables. Indeed,
$\var(t)\subseteq\bigcup_i\var(t_i)$ for deduction rules of the form
\[
\prooftree
S\vdash t_1 \quad \dots \quad S\vdash t_k
\justifies
S\vdash t
\endprooftree
\]
with $k>0$, and $\var(t)\subseteq\var(S)$ for the axiom (that is, if $t\in S$).
\end{proof}

The next lemma shows the ``cut elimination'' property for the deduction system $\vdash$.
\begin{lemma}\label{lemma_deduc-cutelim}
If $T\vdash u$ and $T,u\vdash v$ then $T\vdash v$.
\end{lemma}
\begin{proof}
Consider a proof $\pi$ of $T\vdash u$ and a proof $\pi'$ of $T,u\vdash v$. The tree obtained from~$\pi'$ by
\begin{itemize}
\item
replacing the nodes $T,u\vdash t$ in $\pi'$ with $T\vdash t$,
\item
replacing each new leaf $T\vdash u$ (the old $T,u\vdash u$) with the tree $\pi$,
\end{itemize}
is a proof of $T\vdash v$.
\end{proof}

As a consequence,  if $T\subseteq T'$, $T'\vdash v$, and $T\vdash u$, for all $u\in T'\setminus T$, then $T\vdash v$.

\smallskip

We show now that the simplification rules preserve \dedsyss{}.
\begin{lemma}\label{lemma_c2c}
The simplification rules transform a \dedsys{} into a \dedsys{}.
\end{lemma}
\begin{proof}
Let $C$ be a \dedsys{}, $C=\bigwedge_i (T_i\Vdash
u_i)$ and $C\simpl_\sigma C'$. Since $T_i\subseteq T_{i+1}$
implies $T_i\sigma\subseteq T_{i+1}\sigma$,  $C'$
satisfies the first point of the definition of \dedsyss{}.


We show that $C'$ also satisfies the second point of the definition
of \dedsyss{}. Let $(T'\Vdash u')\in C'$ and $x\in\var(T')$. We have
to prove that $T'_x$ exists and $T'_x\subsetneq T'$. We distinguish cases,
depending on
which simplification rule is applied:

\begin{itemize}
\item 
If the rule $R_1$ is applied, eliminating the constraint
$T\Vdash u$. Then $C'=C\setminus\set{T\Vdash u}$. If $T_x\neq T$ then
$T'_x=T_x$ (and thus $T'_x$ exists and $T'_x\subsetneq T'$). Suppose
that $T_x=T$. Then there is $(T\Vdash u'')\in C$ such that
$x\in\var(u'')$. If $u\neq u''$ then again $T'_x=T_x$ (since
$(T'_x\Vdash u'')\in C'$).  Finally, suppose that $u=u''$. By the
minimality of $T$, it follows that $x\notin\var(T)$ and $x\notin\{y\mid
(T''\Vdash y)\in C, T''\subsetneq T\}$.  Since $x\in\var(u)$, by Lemma~\ref{lemma_deduc-var},
 $T\cup \{y\mid (T''\Vdash y) \in C, T''\subsetneq T\}\not\vdash
u$, which contradicts the applicability of rule $R_1$.

\item If one of the rules $R_2$, $R_3$ or $R'_3$ is applied, then,
 for each constraint $(T''\Vdash u'')\in C'$,
there is a constraint $(T\Vdash u)\in C$ such that $T\sigma=T''$ and $u\sigma=u''$.
Consider $(T\Vdash u)\in C$ such that $T\sigma=T'$ and $u\sigma=u'$.

If $x$ is not introduced by $\sigma$, then $x\in \var(T)$. 
Then $T_x$ exists and $T_x\subsetneq T$. Thus
$T_x\sigma\subseteq T\sigma$. If $T_x\sigma= T\sigma$, then $x\in\var(T_x)$, 
which contradicts the minimality
of $T_x$. Thus $T_x\sigma\subsetneq T\sigma$. We also have that 
$\set{T''\sigma\mid (T''\Vdash u'')\in C,
x\in\var(u'')} \subseteq$ $\{T''\sigma\mid (T''\sigma\Vdash u''\sigma)\in C', x\in\var(u''\sigma)\}$, since,
 for any term $u''$, if $x\in \var(u'')$, then $x\in \var(u''\sigma)$. It follows that $T'_x$ exists and
$T'_x\subseteq T_x\sigma$. Hence $T'_x\subsetneq T'$.


Otherwise, assume that $x$ is introduced by $\sigma$: $\exists
y\in\var(T)$ such that $x\in\var(y\sigma)$. Then $T_y$ exists and
$T_y\subsetneq T$. 
Let $Y=\{z\in\var(T) \mid x\in\var(z\sigma)\}$ and 
let $y_0\in Y$ be such that $T_{y_0}=\min\{T_y \mid y\in Y\}$.
For all $y'\in Y$, we have that 
\begin{align*}
A & \eqdef \{T''\sigma \mid (T''\Vdash u'') \in C', x\in\var(u'')\}  \\
  & = \{T\sigma \mid (T\Vdash u) \in C, x\in\var(u\sigma)\}  \\
  & \supseteq \{T\sigma \mid (T\Vdash u) \in C, \exists z\in\var(u), x\in\var(z\sigma)\}  \\
  & \supseteq \{T\sigma \mid (T\Vdash u) \in C, y'\in\var(u), x\in\var(y'\sigma)\}  \\
  & = \{T\sigma \mid (T\Vdash u) \in C, y'\in\var(u)\} \eqdef B_{y'}.
\end{align*}

Thus $T'_x=\min A \subseteq \min B_{y'}=T_{y'}\sigma$.  From
$T_{y_0}\subsetneq T$, we obtain that $T_{y_0}\sigma\subseteq
T\sigma$. Suppose, by contradiction, that $T_{y_0}\sigma = T\sigma$. Then
$x\in\var(T_{y_0}\sigma)$ (since $x\in\var(T\sigma)$). That is, there
exists $z\in\var(T_{y_0})$ such that $x\in\var(z\sigma)$. From condition 2
of Definition~\ref{def:constraint_sys} applied to $z$, it follows that
$T_z\subsetneq T_{y_0}$. As $z$ is in $Y$, this contradicts the choice
of $y_0$.  Thus $T'_x\subseteq T_{y_0}\sigma\subsetneq T\sigma=T'$.

\item If the rule $R_4$ is applied then there is nothing to prove.

\item If some rule $R_f$ is applied, then the property is preserved, since, if $x\in \var(u'')$ for some term
$u''$ such that $(T''\Vdash u'')\in C'$, then there is a term $v$ with $x\in\var(v)$ such that $(T''\Vdash v)\in C$.
\end{itemize}
\end{proof}

\begin{lemma}[correctness]\label{lemma_correct}\label{lemma:correctness}
If $C\simpl_\sigma C'$, then for every solution $\tau$ for
$C'$, $\sigma\tau$ is a solution of $C$.
\end{lemma}
\begin{proof}
If $C'$ is obtained by applying $R_1$, then we have to prove that $T\tau \vdash u\tau$, where $T\Vdash u$ is the
eliminated constraint.
We know that $T\cup\set{x\mid (T'\Vdash x)\in C, T'\subsetneq T}\vdash u$. It
follows that $T\tau\cup\set{x\tau\mid (T'\Vdash x)\in C, T'\subsetneq T}\vdash u\tau$. 
All constraints $T'\Vdash x$ in $C$ with $T'\subsetneq T$ are also constraints in $C'$.
Thus, for all such constraints, we have that $T'\tau\vdash x\tau$, and hence $T\tau\vdash x\tau$.
Then, from Lemma~\ref{lemma_deduc-cutelim}, we obtain that $T\tau \vdash u\tau$.

If $C'$ is obtained by applying  $
R_2,R_3$ or
$R'_3$, then, for every constraint $T\Vdash u$ of $C$,
$(T\sigma)\tau\vdash (u\sigma)\tau$, hence $T(\sigma\tau)\vdash
u(\sigma\tau)$. 

If $C'$ is obtained by applying some rule $R_f$, then we obtain that $T\tau
\vdash f(u,v)\tau$ from $T\tau \vdash u\tau$ and $T\tau\vdash v\tau$
by applying the corresponding inference rule (e.g. encryption if $f=\enc$).

Finally, $C'$ cannot be obtained by the rule $R_4$, since it is satisfiable.

It follows that, in all cases,
 $\sigma\tau$ satisfies $C$. 
\end{proof}

\subsection{Completeness}\label{app:complet}

Let $T_1\subseteq T_2\subseteq \dots \subseteq T_n$. We say that a proof $\pi$ of $T_i\vdash u$ is
\emph{left minimal} if, whenever
there is a proof of $T_j \vdash u$ for some $j < i$, then, replacing
$T_i$ with $T_j$ in all left members of the labels of $\pi$, yields
a proof of $T_j\vdash u$. In other words, the left-minimal proofs are 
those that can be performed in a minimal $T_j$.

We
say that a proof is \emph{simple} if all its subproofs are left minimal and 
there is no repeated label on any branch.
 Remark that a subproof of a simple proof is simple.

\begin{lemma}\label{lemma_leftminproof}
If there is a proof of $T_i\vdash u$, then there is a simple proof of it.
\end{lemma}
\begin{proof}
We prove the property by induction on the pair $(i,m)$ (considering
 the lexicographic ordering), where $m$ is the size of a proof
of $T_i\vdash u$.

If $i=1$ then any (subproof of any) proof of $T_1\vdash u$ is left minimal and there exists a proof
without repeated labels on any path.

If $i>1$ and there is a $j<i$ such that $T_j\vdash u$, then we apply the
induction hypothesis to obtain the existence of a simple proof of
$T_j\vdash u$. This proof is also a simple proof of $T_i\vdash u$.

If $i>1$ and there is no $j<i$ such that $T_j\vdash u$, then we apply the
induction hypothesis on the immediate subproofs $\pi_1,\ldots,\pi_n$ of the proof $\pi$ of
$T_i\vdash u$. If the label $T_i\vdash u$ appears in one of the
resulting proofs $\pi'_i$, then replace $\pi$ with a subproof of $\pi'_i$ whose
conclusion is $T_i\vdash u$. The new proof does not contain any label  
$T_i\vdash u$.
Otherwise, if $\pi$ is obtained by applying an inference rule $R$ to $\pi_1,\ldots,\pi_n$, then replace $\pi$ with the proof obtained by applying $R$ to
$\pi'_1,\ldots,\pi'_n$.
In both cases the resulting proof and all of its subproofs are left
minimal by construction, and hence the resulting proof is simple.
\end{proof}

\begin{lemma}\label{lemma_axdec}
Let $C$ be a \dedsys{}, $\theta$ be a 
solution of $C$, $T_i$ be a left hand side of $C$ such that, for any
$(T\Vdash v)\in C$, if $T\subsetneq T_i$, then $v$ is a variable. Let
$u$ be any term. If there is a simple proof of $T_i\theta\vdash u$, whose
last inference rule is a decomposition, then there is a non-variable
$t\in\st(T_i)$ such that $t\theta=u$. 
\end{lemma}
\begin{proof}



Consider a simple proof $\pi$ of $T_i\theta\vdash u$. We may assume,
without loss of generality, that $i$ is minimal. Otherwise, we
simply replace everywhere in the proof $T_i$ with a minimal $T_j$ such that 
$T_j\theta\vdash u$ is derivable; by left minimality, we get again a proof tree,
whose last inference rule is a decomposition. Such a $T_j\subseteq T_i$ also
satisfies the hypotheses of the lemma. 
%

We reason by induction on the depth
of the proof $\pi$. We make a case distinction, depending on the last rule of $\pi$:

\begin{description}
\item[The last rule is an axiom] 
Then $u\in T_i\theta$ and there is $t\in T_i$ (thus $t\in\st(T_i)$)
such that $t\theta=u$. By contradiction, if $t$ was a variable then $T_t\Vdash w$, with $t\in \var(w)$ is a
constraint in $C$ such that $T_t\subsetneq T_i$. Moreover, by hypothesis of
the lemma, $w$ must be a variable. Hence $w=t$. Then $T_t\theta \vdash u$,
which contradicts the minimality of $i$.

\item[The last rule is a symmetric decryption]
\[ \pi= \prooftree
\begin{array}{c}\pi_1\\ T_i\theta\vdash \penc{u}{w}\end{array}\;
\;\begin{array}{c}\pi_2\\ T_i\theta\vdash w \end{array}
\justifies T_i\theta\vdash u
\endprooftree
\]

By simplicity, the last rule of $\pi_1$ cannot be a composition:
$T_i\theta\vdash u$ would appear twice on the same path.
Then, by induction hypothesis, there is a non variable $t\in \st(T_i)$
such that $t\theta=\penc{u}{w}$. It follows that
$t=\penc{t'}{t''}$ with $t'\theta=u$. If $t'$ was a
variable, then $T_{t'}\theta\vdash t'\theta$ would be derivable. 
Hence  $T_{t'}\theta\vdash u$ would be derivable, which again contradicts the
minimality of $i$. Hence $t'$ is not variable, 
as required.
\item[The last rule is an asymmetric decryption, (resp. projection, resp. unsigning)]
The proof is similar to the above one:
 by simplicity and by induction hypothesis, there is a non-variable
$t\in \st(T_i)$ such that $t\theta = \enca(u,\pub{v})$ (resp. $t\theta=\pair{u}{v}$, resp. $t\theta=\sign(u,\priv(v))$). 
Then $t= \enca(t',t'')$ (resp. $t=\pair{t'}{t''}$, resp. $t= \sign(t,t'')$).
$t'\in \st(T_i)$, $t'\theta=u$ and, by minimality of $i$, $t'$ is not a variable.
\end{description}
\end{proof}


\begin{lemma}\label{lemma_R1}
Let $C$ be a \dedsys{} and $\theta$ be a solution of $C$. 
Let $T_i$
be a left hand side of a constraint in $C$ and $u$ be a term, such that:
\begin{enumerate}
\item \label{hyp1} for any $(T\Vdash v)\in C$,
if $T\subsetneq T_i$, then $v$ is a variable;
\item \label{hyp2}
 $T_i$ does not contain two distinct non-variable subterms $t_1, t_2$ with
$t_1\theta=t_2\theta$;

\item \label{hyp3} $T_i$ does not contain two terms $\enca(t_1,x)$ and $\priv(t_2)$ where
$x$ is a variable distinct from $t_2$;
\item \label{hyp3b} $T_i$ does not contain two terms $\enca(t_1,t_2)$ and $\priv(x)$ where $x$ is a variable distinct from $t_2$;
\item \label{hyp4} $u$ is a non-variable subterm of $T_i$;
\item \label{hyp5} $T_i\theta \vdash u\theta$.
\end{enumerate}
Then 
 $T'_i\vdash u$, where $T'_i=T_i\cup\set{x\mid (T\Vdash x)\in C, T\subsetneq T_i}$.
\end{lemma}
\begin{proof}
Let $j$ be minimal such that $T_j\theta\vdash u\theta$. Thus $j\le i$ and $T_j\subseteq T_i$.
Consider a simple proof $\pi$ of $T_j\theta\vdash u\theta$.
We reason by induction on the depth of $\pi$. We analyze the different
cases, depending on the last rule of $\pi$:
\begin{description}
\item[The last rule is an axiom]
Suppose, by contradiction, that $u\notin T_j$.
Then there is $t\in T_j$ such that $t\theta=u\theta$ and $t\neq u$.
By hypothesis~\ref{hyp4}, $u$ is not a variable and, by hypothesis~\ref{hyp2} of the lemma,
 $t,u$ cannot be both
non-variable subterms of $T_i$. It follows that $t$ is a variable.
Then $T_t\theta\vdash t\theta$, which implies $T_t\theta\vdash u\theta$,
contradicting the minimality of $j$, since 
$T_t\subsetneq T_j$. 
Hence $u\in T_j$ and then $T'_i\vdash u$, as required.

\item[The last rule is the symmetric decryption rule]
There is $w$ such that $T_j\theta\vdash \penc{u\theta}{w}$,
$T_j\theta\vdash w$:
\[
\prooftree
T_j\theta \vdash \penc{u\theta}{w}\;\;
\; T_j\theta\vdash w 
\justifies T_j\theta\vdash u\theta
\endprooftree
\]
By simplicity, 
the last rule of the proof of $T_j\theta\vdash
\penc{u\theta}{w}$ is a decomposition. 
By Lemma~\ref{lemma_axdec}, there is $t\in\st(T_j)$, $t$ not a variable, 
such that $t\theta=\penc{u\theta}{w}$. Let $t=\penc{t_1}{t_2}$ and $t_1\theta=u\theta$, $t_2\theta=w$.  By induction hypothesis, $T'_i \vdash t$.

If $t_1$ was a variable, then $T_{t_1}\subsetneq T_j$ and, by hypothesis~\ref{hyp1} of
the lemma,
$T_{t_1}$ must be the left-hand-side of
a solved constraint: $(T_{t_1}\Vdash t_1)\in C$ and therefore
 $T_{t_1}\theta \vdash u\theta$,
contradicting the minimality of $j$. 

Now, by hypothesis~\ref{hyp4} of the lemma, $u$ is a non-variable
subterm of $T_i$, hence  $t_1,u$ are two non variable
subterms of $T_i$ such that $t_1\theta=u\theta$. By hypothesis~\ref{hyp2} 
of the lemma,
this implies $t_1=u$. 

On the other hand, if $t_2$ is a variable, 
 $t_2\in \var(T_i)$ implies $T_{t_2}\subsetneq T_i$ and, since $T_i$ is 
minimal unsolved, $(T_{t_2}\Vdash t_2)\in C$, which
implies $t_2\in T'_i$. If $t_2$ is not a variable, then, from $T_j\theta\vdash t_2\theta$ and by induction hypothesis,
$T'_i \vdash t_2$. So, in any case, $T'_i\vdash t_2$.

Now, we have 
 both $T'_i \vdash \enc(u,t_2)$  and $T'_i \vdash t_2$, from which
we conclude that $T'_i \vdash u$, by symmetric decryption.

\item[The last rule is an asymmetric decryption rule]
There is a $w$ such that  $T_j\theta \vdash \priv(w)$ and $T_j\theta \vdash \enca(u\theta,\pub{w})$.  As in the previous case, there is a non-variable
$t\in \st(T_j)$ such that $t\theta = \enca(u\theta,\pub{w})$. 
By induction hypothesis, $T'_i \vdash t$.
Let $t = \enca(t_1,t_2)$.

As in the previous case, $t_1$ cannot be a variable. Therefore $t_1,u$ are
two non-variable subterms of $T_i$ such that $t_1\theta = u\theta$, which
implies that $t_1=u$. (We use here the hypotheses~\ref{hyp2} and~\ref{hyp4}). 

On the other hand, the
last rule in the proof of $T_j\theta \vdash \priv(w)$ is a decomposition
(no composition rule can yield a term headed with $\priv$). Then, by
Lemma~\ref{lemma_axdec} ($T_j$ satisfies the hypotheses of the lemma since
$T_j\subseteq T_i$),  there is a non-variable subterm $w_1\in \st(T_j)$ such
that $w_1\theta = \priv(w)$. Let $w_1=\priv(w_2)$.
By induction hypothesis, $T'_j \vdash \priv(w_2)$.
\[
\prooftree
\begin{array}{rc}
& \enca(t_1,t_2)\theta\\
& \|\\
T_j\theta \vdash & \enca(u\theta,w)
\end{array}\;\;
\;\begin{array}{rc}
& \priv(w_2)\theta\\
& \| \\
T_j\theta \vdash & \priv(w)
\end{array}
\justifies T_j\theta \vdash u\theta
\endprooftree
\]

By hypothesis~\ref{hyp2} of the lemma, $t_2$ and $w_2$ cannot be both non-variable, unless they are identical. 
Then, by hypotheses~\ref{hyp3} and~\ref{hyp3b} of the lemma, 
we must have $t_2=w_2$. Finally, from $T'_i \vdash \enca(u,t_2), T'_i \vdash \priv(t_2)$ we conclude $T'_i \vdash u$. 

\item[The last rule is a projection rule]
\[
\prooftree
T_j\theta\vdash \pair{u\theta}{v}
\justifies T_j\theta \vdash  u\theta
\endprooftree
\]
As before, by simplicity,
the last rule of the proof of $T_j\theta \vdash \pair{u\theta}{v}$
must be a decomposition and, by Lemma~\ref{lemma_axdec}, there is a non variable
term
$t\in \st(T_j)$ such that $t\theta= \pair{u\theta}{v}$. We let $t = \pair{t_1}{t_2}$. By induction hypothesis, $T'_i \vdash t$.

Now, as in the previous cases, $t_1$ cannot be a variable, by minimality of
$T_j$ and hypothesis~\ref{hyp1} of the lemma. Next, by hypotheses~\ref{hyp2}
and~\ref{hyp4}, we must have $t_1=u$. Finally, from $T'_i \vdash \pair{u}{t_2}$
we conclude $T'_i \vdash u$ by projection.

\item[The last rule is an unsigning rule]
\[ \prooftree
T_j\theta\vdash \sign(u\theta,v)
\justifies T_j\theta \vdash u\theta
\endprooftree
\]
This case is identical to the previous one.

\item[The last rule is a composition]
Assume for example that it is the symmetric encryption rule. 
\[
\prooftree
T_j\theta \vdash v_1\;\;
\;T_j\theta \vdash v_2
\justifies T_j\theta \vdash \penc{v_1}{v_2}
\endprooftree
\]
with
$u\theta = \penc{v_1}{v_2}$.
 Since $u$ is not a variable,
$u=\penc{u_1}{u_2}$, $u_1\theta=v_1$, and $u_2\theta=v_2$. If $u_1$
(resp. $u_2$) is a variable then $u_1$ (resp. $u_2$) belongs to
$\var(T_i)$ since $u\in \st(T_i)$. By 
point~2 of Definition~\ref{def:constraint_sys} and hypothesis~\ref{hyp1}
of the lemma, $u_1\in T'_i$ (resp. $u_2\in T'_i$). 

Otherwise, $u_1$ and $u_2$ are
non-variables. Then, by induction hypothesis, $T'_i \vdash u_1$ and $T'_i \vdash u_2$. Hence in both cases
we have $T'_i\vdash u_1$ and $T'_i\vdash u_2$. Thus $T'_i\vdash u$.

The proof is similar for other composition rules. 
\end{description}
\end{proof}

\begin{lemma}[completeness]\label{lemma_complet}\label{lemma:completeness}
If $C$ is an unsolved \dedsys{} and $\theta$ is a solution
of $C$, then there is a \dedsys{} $C'$, a substitution $\sigma$,
and a solution $\tau$ of $C'$ such that
$C\simpl_\sigma C'$ and $\theta=\sigma\tau$.
\end{lemma}
\begin{proof}
Consider a constraint $T_i\Vdash u_i$ such that, for any $(T\Vdash v) \in C$ such that
 $T\subsetneq T_i$, $v$ is a variable and assume $u_i$ is not a variable. 
If $C$ is unsolved, there is such a constraint in $C$. 

Since $\theta$ is a solution, $T_i\theta\vdash u_i\theta$. 
Consider a simple proof of
$T_i\theta\vdash u_i\theta$. We distinguish cases, depending on the last
rule applied in this proof:
\begin{description}
\item[The last rule is a composition] Since $u$ is not
a variable, $u=f(u_1,\ldots,u_n)$ and $T_i\theta \vdash u_j\theta$ for
every $j=1,...,n$.  Then we may apply the transformation rule
$R_f$ to $C$, yielding constraints $T_i\Vdash u_j$ in $C'$ for every $j$.
$\theta$ is a solution of the resulting \dedsys{} $C'$ by hypothesis.




\item[The last rule is an axiom or a decomposition] 
By Lemma~\ref{lemma_axdec}, there is a non-variable term $t\in \st(T_i)$
such that $t\theta=u_i\theta$.
We distinguish then again between cases, depending on $t,u_i$:

\begin{description}
\item[Case $t\neq u_i$] Then, since $t,u_i$ are both non-variable terms,
 we may apply the simplification rule $R_2$ to $C$: $C \simpl_\sigma C'$
where $C'=C\sigma$ and $\sigma = \mgu(t,u_i)$. Furthermore, $t\theta = u_i\theta$, hence (by definition of a mgu) there is a substitution $\tau$ such that
$\theta = \sigma\tau$. Finally, $\theta$ is a solution of $C$, hence $\tau$ is
a solution of $C'$.
\item[Case $t=u_i$] Then $u_i\in \st(T_i)$. 
\begin{enumerate}
\item\label{case_R3} If there are two distinct non-variable terms $t_1, t_2\in \st(T_i)$ such that
$t_1\theta=t_2\theta$. Then we
apply the simplification rule $R_3$, yielding a \dedsys{} $C'=C\sigma$.
As in the previous case, there is a substitution $\tau$ such that $\theta=\sigma\tau$ and $\tau$ is a solution of $C'$.
\item If there are $\enca(t_1,t_2), \priv(t_3)\in \st(T_i)$ such that
either $t_2$ or $t_3$ is a variable, $t_2\neq t_3$ and $t_2\theta=t_3\theta$,
then we may apply the rule $R'_3$ and conclude as in the previous case.
\item Otherwise, 
we match all hypotheses of Lemma~\ref{lemma_R1} and we conclude
that $T'_i \vdash u_i$. Then the rule $R_1$ can be applied to $C$,
yielding a \dedsys{}, of which $\theta$ is again a solution.
\end{enumerate}
\end{description}
\end{description}
\end{proof}

\subsection{Termination}
\label{section:termination}

The simplification rules also terminate, whatever strategy is used for
their application:
\begin{lemma}\label{lemma:termination}
The constraint simplification rules of Figure~\ref{fig:rules} are (strongly)
terminating.
\end{lemma}

\begin{proof}
Interpret any \dedsys{} $C$ as a pair of non-negative integers
$I(C)=(n,m)$ where $n$ is the number of variables of the system and $m$ is the
number of function symbols occurring in the right hand sides of the system
(here, we assume no sharing of subterms). 
If $C \simpl_\sigma C'$, then $I(C) >_{lex} I(C')$ where $\geq_{lex}$ is the lexicographic ordering
on pairs of integers. Indeed, the first component strictly decreases by 
applying $R_2,R_3,R'_3$, and any other rule strictly decreases the second
component, while not increasing the first one.
The well foundedness of the lexicographic extension of a well-founded
ordering implies the termination of any sequence of rules. 
\end{proof}

\subsection{Proof of Theorem~\ref{theo:general}}
\label{section:th-proof}

Theorem~\ref{theo:general} follows from Lemmas~\ref{lemma:correctness},
\ref{lemma:completeness}, and~\ref{lemma:termination}, by induction on the derivation length, and since \dedsyss{} on
which no simplification rule can be applied must be solved. 
Note that the extension of the correctness and completness lemmas to
security properties is trivial. Indeed, if $\phi$ is a (in)security
property, then $\theta$ is a solution of $\phi\sigma$ if and only if
$\sigma\theta$ is a solution of $\phi$, for any substitutions $\theta$
and $\sigma$.

\subsection{A decision procedure in NP-time}\label{sec:poly}
\label{section:complexity}

The termination proof of the last section does not provide with tight complexity bounds.
In fact, applying the simplification rules may lead to
branches of exponential length (in the size of the constraint system).
Indeed when applying a simplification
rule to a \dedcons{}, the initial constraint is removed from the
constraint system and replaced by new constraint(s). But this
\dedcons{} may appear again later on, due to other simplification
rules. It is the case for example when considering the
following \dedsys{}.
\begin{eqnarray*}
T_0 \eqdef \{\enc(a,k_0)\} & \Vdash & \enc(x_0, k_0)\\
T_1 \eqdef T_0\cup\{\enc(\langle x_0,\langle x_0,a\rangle\rangle,k_1)\}
& \Vdash & \enc(x_1,k_1)\\
&\vdots &\\
T_n \eqdef T_{n-1}\cup\{\enc(\langle x_{n-1},\langle
x_{n-1},a\rangle\rangle,k_n)\}
& \Vdash & \enc(x_n,k_n)\\
T_{n+1} \eqdef T_n\cup\{a\} & \Vdash & x_n
\end{eqnarray*}

The \dedsys{} $C$ is clearly satisfiable
and its size is linear in $n$.
We have that
\[
C\simpl^{2n}_{\sigma}
\left\{\begin{array}{rcl}
T_0&\Vdash &\enc(x_0, k_0)\\
T_{n+1}\sigma & \Vdash & x_n\sigma\end{array}\right.
\]
with $\sigma(x_{i+1}) = \langle x_i,\langle x_i,a\rangle\rangle$ for
$0\leq i\leq n-1$. This derivation is obtained by applying rule $R_2$ and then $R_1$ for each constraint
$T_i\Vdash \enc(x_i,k_i)$ with $1\leq i\leq n$.
The rule $R_1$ cannot be applied to $T_{n+1}\sigma  \Vdash  x_n\sigma$
since $x_0$ and the keys $k_i$ are not present in or derivable from $T_{n+1}\sigma$.
Note that $\sigma'= \sigma\cup \{\subst{x_0}{a}\}$ is a
solution of $C$ and can be easily obtained by rule $R_2$ on the first constraint and then rule $R_1$ on both
constraints.

However, there is a branch of length $3(2^n-1)$ from $T\Vdash x_n\sigma$ leading to $T\Vdash x_0$ (in solved
form), where $T$ denotes $T_{n+1}\sigma$.
This is easy to see by induction on $n$.
It is true for $n = 0$. Then using only the rules
$R_{\langle\,\rangle}$ and $R_1$, we have
\begin{multline*}
T\Vdash x_{n}\sigma\stackrel{R_{\langle\rangle}}{\simpl}
\left\{\begin{array}{rcl}
T&\Vdash& x_{n-1}\sigma\\
T&\Vdash &\langle x_{n-1}\sigma,a\rangle
       \end{array}\right.
\simpl^m
\left\{\begin{array}{rcl}
T&\Vdash& x_{0}\\
T&\Vdash& \langle x_{n-1}\sigma,a\rangle
       \end{array}\right.
\stackrel{R_{\langle\rangle}}{\simpl}\\
\left\{\begin{array}{rcl}
T&\Vdash &x_{0}\\
T&\Vdash &x_{n-1}\sigma\\
T&\Vdash &a
       \end{array}\right.
\stackrel{R_{1}}{\simpl}
\left\{\begin{array}{rcl}
T&\Vdash &x_{0}\\
T&\Vdash &x_{n-1}\sigma
       \end{array}\right.
\simpl^m T\Vdash x_{0}
\end{multline*}
with $m= 3(2^{n-1}-1)$ by induction hypothesis. The length of the branch is
$2\times 3(2^{n-1}-1)+3= 3(2^{n}-1)$.
This shows that there exist branches of exponential length in the size
of the constraint.



\bigskip

We can prove 
that it is
actually not useful to consider \dedconss{} that have already been
seen before (like the constraint $T\Vdash x_{n-1}\sigma$ in our
example). Thus we memorize the constraints that have already been
visited. The constraint simplification rules, instead of operating
on a single \dedsys{}, rewrite a pair of two constraint systems,
the second one representing \dedconss{} that have already been 
processed at this stage: if $C\simpl_\sigma C'$, then
$$C;D\ \msimpl_{\sigma}\ C'\setminus D; D\cup(C\setminus C')$$ 
The constraints (``memorized'') in $D$ are those which were already analyzed (i.e. transformed or eliminated).
The initial constraint system  is $C;\emptyset$.

First, memorization indeed prevents from performing several times the
same transformation:
\begin{lemma}\label{lemma:disjoint}
If $C$ is a \dedsys{} and
$C;\emptyset\msimpl^*_{\sigma} C';D'$ then $C'\cap D'=\emptyset$. 
\end{lemma}

\begin{proof}
\[(C'\setminus D) \cap ((C\setminus C') \cup D)= ((C'\setminus D) \cap D )
\cup ((C'\setminus D) \cap (C\setminus C'))=\emptyset
\]
\end{proof}

This kind of memorization is correct and complete in a more general
setting. We assume in this section that the reader is familiar with
the usual notions of first-order formulas, first-order structures, and
models of first-order logic.

A \emph{(general) constraint} is a (first-order) formula, together
with an interpretation structure~$S$. A \emph{(general) constraint system}
$C$ is a finite set of constraints, whose interpretation is the same as
their conjunction. If $\sigma$ is an assignment of the free variables of
$C$ to the domain of $S$, $\sigma$ is a \emph{solution} of $C$ if $\sigma, S \models
C$. In the context of constraint systems, $S$ is omitted: the
satisfaction relation $\models$ refers implicitly to $S$. It is extended,
as usual, to entailment: $C \models C'$ if any solution of $C$ is also a
solution of $C'$.
%
%
We may consider constraints $c$ as singleton constraint systems, and
thus write for example $c \models c'$ instead of $\{c\} \models
\{c'\}$.

A \emph{(general) constraint system transformation} is a binary
relation $\gsimpl$ on constraints such that, for any sequence (finite
or infinite) $C_1 \gsimpl \cdots \gsimpl C_n \gsimpl \cdots$, there is
an ordering $\geq$ on individual constraints such that, for every $i$,
for every $c\in C_{i}\setminus C_{i+1}$, we have 
\begin{equation}\label{gtransf}
\{d\in C_{i+1} \mid d < c\} \models c.
\end{equation}
This expresses the \emph{correctness} of the
transformations: only redundant formulas are removed. The ordering
needs not to be well-founded.

Our \dedsyss{} and \dedcons{} simplification rules satisfy these
properties.  More precisely, we need to consider the substitutions
(partial assignments) as part of the constraint system, in order to
fit with the above definition: constraint systems come in two parts: a
set of deducibility constraints and a set of solved equations,
recording the substitution computed so-far.
In other words, a sequence of simplification steps
$C_0\rightsquigarrow_{\sigma_1} C_1 \rightsquigarrow_{\sigma_2} \dots$
can be written as a general transformation sequence $C_0 \leadsto (C_1
\wedge \sigma_1) \leadsto (C_2 \wedge \sigma_1 \wedge \sigma_2)
\leadsto \dots$, where substitutions $\{\subst{x_1}{t_1},\dots,
\subst{x_n}{t_n}\}$ are seen as conjunctions of solved equations
$(x_1=t_1)\wedge \dots \wedge (x_n=t_n)$.

We show next that for any sequence $C_0\rightsquigarrow_{\sigma_1} C_1
\rightsquigarrow_{\sigma_2} \dots$ of simplification steps there is an
ordering $\ge$ on the corresponding general constraints such that
(\ref{gtransf}) holds.

We start by defining the ordering.
First, we order the variables by $x>y$ if, for some~$i$,
$y\in\var(x\sigma_1\dots\sigma_i)$. Intuitively, $x>y$ if $x$ is
instantiated before $y$ in the considered derivation. Indeed, let
$i_x$ be the minimum among all indexes $i$ such that $x\sigma_i\neq x$
if this minimum exists and $\infty$ otherwise. Then $x>y$ implies that
either $i_x<i_y$, or $i_x=i_y$ and $y\in \var(x\sigma_{i_x})$. (Note
that in this last case we cannot have both $y\in \var(x\sigma_{i_x})$
and $x\in \var(y\sigma_{i_x})$, by the definition of a $\mgu$.) This
observation proves that the relation $>$ on variables is an ordering.
Next, we let $(T \Vdash u) > (T'\Vdash u')$ if
\begin{itemize}
\item either the multiset of variables occurring in $T$ is
strictly larger than the multiset of variables occurring in $T'$;
such multisets are ordered by the multiset extension of the ordering on
variables;
\item or else the multisets of variables are identical, and $T'\subsetneq T$;
\item or else $T=T'$ and the multiset of variables in $u$ is strictly
larger than the multiset of variables in $u'$;
\item or else, $T=T'$, the multisets of variable are identical
 and the size of $u$ is strictly larger than the
size of $u'$.
\end{itemize}  
This is an ordering as a lexicographic composition of orderings.
Finally, any solved equation (i.e.~substitution) is strictly smaller
than any deducibility constraint, and equations are not comparable.

The ordering we have just defined could have been used for the
termination proof, as it is a well-founded ordering.  It will now be
considered as the default ordering on constraints, when a derivation
sequence is fixed.

This ordering also satisfies the above required hypotheses for general
constraint system transformations, as shown by the proof of the
following proposition.

\begin{proposition}\label{prop-isgtransf}
  The simplification rules on \dedsyss{} form a general constraint
  system transformation.
\end{proposition}
\begin{proof}
  Let $C_0\rightsquigarrow_{\sigma_0} C_1 \rightsquigarrow_{\sigma_1}
  \dots$ be a simplification sequence.  We consider the ordering on
  \dedconss{} (viewed as general constraints) defined above.

  We show next that (\ref{gtransf}) holds.  Note that in
  (\ref{gtransf}), $c$ cannot be a solved equation, because at each
  step solved equations ($x=x\sigma_i$) may be added but no equation
  is eliminated.
  Thus let $(T\Vdash u)\in C_{i}\setminus C_{i+1}$, for some $i\ge 0$.
  We need to show that 
%
\begin{equation}\label{gtransf-inst}
  \bigwedge_{\substack{(T'\Vdash u')\in C_{i+1}\\(T'\Vdash u') <
      (T\Vdash u)}} T'\Vdash u'\ \wedge\ \bigwedge_{1\le
    j\le i}\sigma_j\quad \models \quad T\Vdash u
\end{equation}
%
  We investigate the possible transformation rules.

  For the rules $R_2,R_3,R'_3$, $C_{i+1}=C_i\sigma_i$. We have
  $(T\Vdash u) \geq (T\sigma_i\Vdash u\sigma_i)$ since either the
  multiset of variables of $T\sigma_i$ is strictly smaller than the
  multiset of variables of $T$, or else $T=T\sigma_i$ and, in the
  latter case, either the multiset of variables of $u\sigma_i$ is
  strictly smaller than the multiset of variables of $u$ or else
  $u\sigma_i=u$.  Moreover, $c\sigma \wedge \sigma \models c$ for all
  constraints $c$ and substitutions~$\sigma$.
  Indeed, if $\theta$ is a solution of $c\sigma \wedge \sigma$ then
  $x\theta=x\sigma\theta$ for any $x\in\dom(\sigma)$. It follows that
  $c\theta=c\sigma\theta$, and thus $\theta$ is a solution of $c$.

  Hence, we have in particular that $(T\sigma_i\Vdash u\sigma_i)\wedge
  \sigma_i\models T\vdash u$, which shows that (\ref{gtransf-inst})
  holds for this case.

  For the rule $R_f$, it suffices to notice that $\{T\Vdash
  u_1,\ldots, T\Vdash u_n\} \models (T\Vdash f(u_1,\ldots,u_n))$ and
  $(T\Vdash u_i) < (T\Vdash f(u_1,\ldots,u_n))$ for every $i$.

For the rule $R_1$, the constraint $T\Vdash u$ is a consequence of the
(strictly smaller) constraints $T' \Vdash x$ for $T'\subsetneq T$. 

Finally, the rule $R_4$ only applies to unsatisfiable \dedconss{}.
\end{proof}

The memorization strategy can be defined, as above,
 for any general constraint system transformation.
 The correctness of the memorization strategy
relies on the following invariant:

\begin{lemma}\label{lemma:redundancy}
For any  constraint system transformation $\gsimpl$, if $C;\emptyset \leadsto^* C'; D'$, then $C' \models D'$.
\end{lemma}

\begin{proof}
We prove, by induction on the length of the derivation sequence the following
stronger  result: $\forall d\in D',  \{c\in C'\mid c < d\} \models d$.

The base case is straightforward as $D'$ is empty. Next, assume that
$C; D \leadsto C'; D'$. By definition, $D'= D \cup (C \setminus C')$. If
$d\in C\setminus C'$, by definition of a constraint transformation rule,
$\{c\in C'\mid c< d\} \models d$. If $d\in D$, by induction hypothesis,
$\{c\in C\mid c < d\} \models d$. Hence $\{c\in C'\: | c < d\} \cup \{c\in C\setminus C' \mid c< d\} \models d$. But, again by definition of constraint transformations, any constraint in the second set is a consequence of the first set:
we get $\{c\in C'\: | c< d\} \models d$. 
\end{proof}

It follows that the memorization strategy is always correct when the
original constraint transformation is correct.

Now, the memorization strategy preserves the properties of our \dedsyss{}:
\begin{lemma}\label{lemma:preserves_origination}
If $C$ is a \dedsys{} and
$C;\emptyset\msimpl^*_{\sigma} C';D'$ then $C'$ is a \dedsys{}.
\end{lemma}
\begin{proof}
Let $(C_i;D_i)\msimpl_{\sigma_{i+1}}(C_{i+1};D_{i+1})$, with $0\le i< n$ be the sequence of
\dedsyss{} obtained by applying
successively the simplification rules, where $C_0=C$,
$D_0=\emptyset$, $C_n=C'$, and
$C_i\msimpl_{\sigma_{i+1}} C'_{i+1}$ (and thus $C_{i+1}=C'_{i+1}\setminus D_i$, and $D_{i+1}=D_i\cup (C_i\setminus C'_{i+1})$).
We know that $C'_i$ is a \dedsys{}, by Lemma~\ref{lemma_c2c}. 

First, the left members of $C_i$ are linearly ordered by inclusion, as
they are a subset of the left members of $C'_i$. 

We consider now the other property of \dedsyss{}.
We let $\geq$ be the ordering on constraints defined before. 
We show below, by induction on $i$ that, for every $x\in \var(C_i)$,
for every $(T\Vdash u)\in D_i$ such that $x\in \var(u)\setminus\var(T)$,
there is a $(T'\Vdash u') \in C_i$ such that $x\in \var(u')\setminus\var(T')$
and $(T'\Vdash u') < (T\Vdash u)$. 

Note that this property implies that $C_i$ is a \dedsys{}: For every
variable $x\in \var(C_i)$, there is $(T_x\Vdash u)\in C'_i$ such that
$x\in \var(u)\setminus \var(T_x)$, as $C'_i$ is a \dedsys{}.  If
$(T_x\Vdash u)\in C_i$ then we're done, otherwise $(T_x\Vdash u)\in
D_i$, and hence, by the stated property, there is $(T'_x\Vdash u')\in
C_i$ such that $x\in \var(u')\setminus \var(T'_x)$. This shows that
$C_i$ is a \dedsys{}.

The property holds trivially for $i=0$. For the induction step,
let  $x\in\var(C_{i+1})$ and  $(T\Vdash u)\in C'_{i+1}$ be such that $x\in\var(u)\setminus \var(T)$.
We investigate three cases:
\begin{itemize}
\item if $C_{i+1}$ is obtained by one of the rules $R_2,R_3,R'_3$, then $C_{i+1}= C_i\sigma_{i+1}\setminus D_i$,
and $x\notin \dom(\sigma_{i+1})$.
We assume w.l.o.g.~that $T\Vdash u$ is a minimal constraint in $D_{i+1}$ such
that $x\in \var(u)\setminus \var(T)$. 

There is $(T'\Vdash u') \in C_i$ such that $x\in \var(u')\setminus \var(T')$ and $(T'\Vdash u') \leq (T\Vdash u)$:
if $(T\Vdash u)\notin C_i$, then $(T\Vdash u) \in D_i$ and
by induction hypothesis, there is a $(T'\Vdash u')\in C_i$ such that 
$x\in \var(u')\setminus \var(T')$ and $(T'\Vdash u') < (T\Vdash u)$.

Let $S = \{y\in \var(T')\mid x\in \var(y\sigma_{i+1})\}$. By induction
hypothesis $C_i$ is a constraint system, and hence, for every $y\in
S$, there is a (minimal) constraint $T_y\Vdash u_y\in C_i$ such that
$y\in \var(u_y)\setminus \var(T_y)$. Since $y\in \var(T')$,
$T_y\subsetneq T'$.  Let $T_1\Vdash u_1$ be a minimal element in
$\{T_y\Vdash u_y\mid y\in S\}\cup \{T'\Vdash u'\}$. 
Suppose that $x\in\var(T_1\sigma_{i+1})$.  Since $x\notin\var(T')$ and
$T_y\subsetneq T'$, it follows that $x\notin \var(T_y)$, and hence
there is $z\in\var(T_1)$ such that $x\in\var(z\sigma_{i+1})$. It
follows that $z\in S$ and $T_z\subsetneq T_1$, which contradicts the
minimality of $T_1\Vdash u_1$. Hence $x\in
\var(u_1\sigma_{i+1})\setminus \var(T_1\sigma_{i+1})$.
Also $(T_1\sigma_{i+1}\Vdash u_1\sigma_{i+1}) \leq (T_1\Vdash u_1) \leq
(T'\Vdash u') \leq (T\Vdash u)$. Furthermore, at least one of the
inequalities is strict: if $(T\Vdash u)\in D_i$ the last inequality is
strict, otherwise $(T\Vdash u)\in (C_i\setminus C'_{i+1}) = (C_i\setminus C_i\sigma)$ hence
$(T\sigma_{i+1} \Vdash u\sigma_{i+1}) < (T\Vdash u)$.  It follows that
$(T_1\sigma_{i+1}\Vdash u_1\sigma_{i+1})\in C_{i+1}$ by minimality of
$T\Vdash u$.

 




\item if $C_{i+1}$ is obtained by an $R_f$ rule. 
We may assume w.l.o.g.~that $T\Vdash u$ is a minimal constraint in $D_{i+1}$ such that $x\in \var(u)\setminus \var(T)$. 

Either $(T\Vdash u)\in D_i$,
in which case, by induction hypothesis, there is $(T'\Vdash u')\in C_i$ such
that $x\in \var(u')\setminus \var(T')$ and $(T'\Vdash u') < (T\Vdash u)$. 
If $(T'\Vdash u')\in C_{i+1}$, there
is nothing to prove. Otherwise, $u'=f(u_1,\ldots,u_n)$ and, for every $j$,
$(T'\Vdash u_j)\in C_{i+1}\cup D_i$. 
Moreover, there is an index $j$ such that
$x\in \var(u_j)\setminus \var(T')$ and, by minimality of $T\Vdash u$, $(T'\Vdash u_j)\in C_{i+1}$, hence completing this case.

Or else $(T\Vdash u) \in C_i\setminus C'_{i+1}$, in which case $u=f(u_1,\ldots,u_n)$ and $(T\Vdash u_j)\in C_{i+1}\cup D_i$. As above, we conclude that for some
$j$, $x\in\var(u_j)\setminus\var(T), (T\Vdash u_j)\in C_{i+1}$ and $(T\Vdash u_j) < (T\Vdash u)$.

\item if $C_{i+1}$ is obtained by the rule $R_1$, removing a constraint $T_1\Vdash u_1$, then $D_{i+1}=D_i\cup\{T_1\Vdash u_1\}$ and, by Lemma~\ref{lemma_c2c} for any variable $y\in \var(u_1)\setminus\var(T_1)$ there is a strictly smaller constraint $(T_2\Vdash u_2)\in C_i$ such that
$y\in\var(u_2)\setminus\var(T_2)$. Then we simply apply the induction hypothesis.
\end{itemize}
\end{proof}

\begin{theorem}\label{theo:general_NP}
Let $C$ be a \dedsys{}, $\theta$ a substitution and  $\phi$ a security property. 
\begin{enumerate}
\item(Correctness)
If $C;\emptyset\msimpl^*_{\sigma} C';D'$ for some \dedsys{}
$C'$
and some substitution $\sigma$, if $\theta$ is an attack  for $C'$ and $\phi\sigma$,
then $\sigma\theta$ is an attack for $C$ and $\phi$.
\item(Completeness) If $\theta$ is an attack for  $C$ and $\phi$,
then there exist a \dedsys{} $C'$ in solved
  form, a set of \dedconss{} $D'$
  and  substitutions $\sigma,\theta'$ such that $\theta
  =\sigma\theta'$, $C;\emptyset\msimpl^*_{\sigma} C';D'$, and $\theta'$ is an
attack for  $C'$ and $\phi\sigma$.
\item(Termination) If $C;\emptyset\msimpl^n_{\sigma} C';D'$ for some
\dedsys{}
$C'$
and some substitution $\sigma$, then $n$ is polynomially bounded in
the size of $C$.
\end{enumerate}
\end{theorem}

\begin{proof}
For correctness, we rely on Lemmas~\ref{lemma_correct},
and~\ref{lemma:redundancy}: by Lemma~\ref{lemma:redundancy}, any solution  $\theta$ of $C'$ 
is also a solution $C' \cup D'\sigma$ and, by Lemma~\ref{lemma_correct} (and induction),
$\sigma\theta$ is a solution of $C$. 

For completeness, 
from Lemma~\ref{lemma_complet}, we know that
if $C_i$ is an unsolved \dedsys{} and $\theta$ is an attack
for $C_i$ and $\phi$, 
then there is a \dedsys{} $C'_{i+1}$, a substitution $\sigma_i$,
and an attack $\tau_i$ for $C'_{i+1}$ and $\phi\sigma_i$ 
such that $C_{i}\simpl_{\sigma_{i}} C'_{i+1}$ and $\theta=\sigma_i\tau_i$.
Then $\tau_i$ is an attack also for $C'_{i+1}\setminus D_i$ and
$\phi\sigma$, for any set of constraints $D_i$. By
Lemma~\ref{lemma:preserves_origination}, we know that when $D_i$
represents already visited constraints, then $C'_{i+1}\setminus D_i$
is a \dedsys{}. We can thus conclude by induction on the derivation
length $n$, taking $C_0=C$, $D_0=\emptyset$, $C_{i+1}=C'_{i+1}\setminus
D_i$ for all $i$, and $C_n=C'$.


Concerning termination, we assume a DAG representation of
the terms and constraints, in such a way that the size of the constraint
is proportional to the number of the distinct subterms occurring in it.
Next, observe that $\card\st(t\sigma) \leq \card(\st(t)\cup \bigcup_{x\in\dom(\theta)} \st(x\theta))$. Hence, when unifying two subterms of $t$, with
$\mgu$ $\theta$,
$\card\st(t\theta) \leq \card\st(t)$ since, for every variable $x\in \dom(\theta)$, $x\theta$ is a subterm of $t$. It follows that, for any constraint system
$C';D'$ such that $C;\emptyset \msimpl_\sigma^* C';D'$, $\card\st(C') \leq \card\st(C)$. 

Next, observe that the number of distinct left hand sides of the constraints
$\card\lhs(C')$ is never increasing: $\card\lhs(C') \leq \card\lhs(C)$.
Furthermore, as long as we only apply the rules $R_1,R_f$, starting from $C''$,
 the 
left hand sides of the \dedsyss{} are fixed: there are at most
$\card\lhs(C'')$ of them. 
Now, since, thanks to memorization, we cannot get twice the same
constraint, the number of consecutive $R_1,R_f$ steps is bounded by
\[\card\lhs(C'') \times \card\st(\rhs(C'')) \leq \card\lhs(C) \times 
\card\st(C) \]

It follows that the length of a derivation sequence is bounded
by  $\card\var(C)\times \card\lhs(C) \times \card\st(C)$ (for $R_1,R_f$ steps)
plus $\card\var(C)$ (for $R_2,R_3,R'_3$ steps) plus $1$ (for a possible
$R_4$ step).
\end{proof}

Theorem~\ref{theo:general_NP} extends the result of~\cite{RT01} to
sorted messages and general security properties.
Handling arbitrary security properties is possible as soon as we
do not forget any solution of the \dedsyss{} (as we do). If we only
preserve the existence of a solution of the constraint (as in \cite{RT01}),
it might be the case that the solution of $C$ that we kept is not a solution
of the property $\phi$, while there are solutions of both $\phi$ and $C$,
that were lost in the satisfiability decision of $C$. 
In addition, compared to~\cite{RT01}, presenting the decision
procedure using a small set of simplification rules makes it more
easily amendable to further extensions and modifications.
For example, Theorem~\ref{theo:general_NP} has been used in~\cite{CKKW-fsttcs2006} for
proving that a new notion of secrecy in presence of hashes is
decidable (and co-NP-complete) for a bounded number of sessions.

Note that termination in polynomial time also requires the use of a
DAG (Directed Acyclic Graph) representation for terms.


The following corollary
is easily obtained from the previous theorem by observing that we
can guess the simplification rules which lead to a solved form.

\begin{corollary}\label{corol:general}
Any property $\phi$ that can be decided
in polynomial time on solved \dedsyss{} can be decided in non-deterministic polynomial
time on arbitrary \dedsyss{}.
\end{corollary}

\subsection{An alternative approach to polynomial-time termination}
\label{section:strategy}

Inspecting the completeness proof, there is still some room for choosing a
strategy, while keeping completeness (correctness is independent
of the order of the rules application). 
To obtain even more flexibility, we slightly relax the condition on
the application of the rule $R_2$ on a constraint $T\Vdash u$: we
require unifying a subterm $t\in\st(T)$ and a subterm $t'\in\st(u)$
(instead of unifying $t$ with $u$) where, as before, $t\neq t'$, $t$,
$t'$ non-variables.  Remark that this change preserves the
completeness of the procedure.

Let us group the rules $R_2,R_3,R'_3$ and call them \emph{substitution
  rules} $S$. We write $S(u,v)$ if the substitution is obtained by
unifying $u$ and $v$.  There are some basic observations:
\begin{enumerate}
\item \label{obs1} 
If $C \simpl^{R_f} C' \simpl_\sigma^S C'\sigma$, then 
$C \simpl^{S}_\sigma C\sigma \simpl^{R_f} C'\sigma$. Hence we may always move
forward the substitution rules.
\item \label{obs2} 
If $C_1\simpl^{R_f} C'_1$ and $C_2 \simpl^{R_f} C'_2$, then
$C_1 \wedge C_2 \simpl^{R_f} C'_1 \wedge C_2 \simpl^{R_f} C'_1\wedge C'_2$ and
$C_1 \wedge C_2 \simpl^{R_f} C_1 \wedge C'_2 \simpl^{R_f} C'_1\wedge C'_2$,
hence any two consecutive applications of $R_f$ on different constraints
can be performed in any order. 
\item \label{obs3} 
The rules $R_1,R_4$ can be applied
at any time when they are enabled; we may apply them eagerly or postpone
them until no other rule can be applied.
\item \label{obs4} 
If $C \simpl^{S(u_1,v_1)}_{\sigma_1} C\sigma_1 \simpl^{S(u_2\sigma_1,v_2\sigma_1)}_{\sigma_2} C\sigma_1\sigma_2$, then, for some
$\theta_1,\theta_2$, 
\[C \simpl^{S(u_2,v_2)}_{\theta_1} C\theta_1 \simpl^{S(u_1\theta_1,v_1\theta_1)}_{\theta_2} C\sigma_1\sigma_2\] Hence any two consecutive
substitution rules can be performed in any order. 
\item \label{obs5} 
If $C\simpl^S_\sigma C\sigma \simpl^{R_f} C'\sigma$, and $S\neq R_2$,
 then $C\simpl^{R_f} C' \simpl^{S}_\sigma C'\sigma$.
\end{enumerate}

This provides with several complete strategies. For instance the
following strategy is complete:
\begin{itemize}
\item apply eagerly $R_4$ and postpone $R_1$ as much as possible
\item apply the substitution rules eagerly (as soon as they are enabled).
This implies that all substitution rules are applied at once, since the
rules $R_1,R_4,R_f$ cannot enable a substitution.
\item when $R_4$ and substitutions rules are not enabled, apply
$R_f$ to the constraint, whose right hand side is maximal (in size).
\end{itemize}
Such a strategy will also yield polynomial length derivations, since
we cannot get twice the same constraint: in any derivation sequence
$C_0\simpl_{\sigma_1} \cdots \simpl_{\sigma_n} C_n$,
if $(T\Vdash u) \in C_i\setminus C_{i+1}$ (we say then that $T\Vdash u$
has been eliminated at this step), then, for any $j >i$, $(T\Vdash u)\notin C_j$.
Indeed, for the substitution rules, $T\Vdash u$ is eliminated only  when $x\in \var(T\Vdash u)$
and $x\in \dom(\sigma_{i+1})$, in which case for any $j>i$, $x\notin\var(C_j)$.
And, if  $T\Vdash u$ is eliminated by an $R_f$ rule, then
 $|u|=\max_{t\in\rhs(C_i)}|t|$. If, for some $j>i$, the
constraint $T\Vdash u$ was in $C_{j+1}$ and not in  $C_j$, 
 then we would have $\max_{t\in\rhs(C_j)}|t|>|u|$.
Thus the maximum of the sizes of the right hand sides terms would have increased, which is not possible according to our strategy.

Then the complexity analysis of
the proof of Theorem~\ref{theo:general_NP} can be applied here.
\\

The above observations can also be used to bound the non-determinism
(which is useful in practice): 
for instance from (\ref{obs1}) and (\ref{obs4}), we see that
substitution rules can be applied ``don't care'': if we use
a substitution rule, we do not need to consider other alternatives.  
More precisely, if $S(t,u)$ is a substitution rule that is applicable to
$C$, let $\Phi(C)$ be the set of substitution rules $S(t',u')$, which
are applicable to $C$ and such that there is no $\theta$ other than the identity
such that
$\mgu(t,u)\theta = \mgu(t',u')$. 
Then
\[ \theta \models C \;\;\Longrightarrow \;\; \displaystyle{\bigvee_{S(t',u')\in \Phi(C)} \exists \theta'. \; \theta= \mgu(t',u')\theta'} \]

Similarly, from (\ref{obs5}), a right-hand side member
that is not unifiable with a non-variable subterm of the corresponding
left hand side, can be ``don't care'' decomposed:
\[ \theta \models C\wedge (T\Vdash f(u_1,\ldots,u_n)) \;\; \Longrightarrow \;\;
\theta \models C\wedge (T\Vdash u_1)\wedge\ldots\wedge (T\Vdash u_n) \]
if $f(u_1,\ldots,u_n)$ is not unifiable with any non-variable subterm of $T$.

\section{Decidability of encryption cycles}\label{sec:cycles}
Using the general approach presented in the
previous section, verifying particular properties like the existence
of key cycles or the conformation to an \emph{a priori} given ordering
relation on keys can be reduced to deciding these properties on solved
\dedsyss{}. We deduce a new decidability result, useful in
models designed for proving cryptographic properties. 

To show that formal models (like the one presented in this article)
are sound with respect to cryptographic ones, the authors usually
assume that no key cycle can be produced during the execution of a
protocol or, even stronger, assume that the ``encrypts'' relation on keys
follows an \emph{a priori} given ordering.

For simplicity, and since there are very few papers constraining the key
relations in an asymmetric setting, in this section we restrict our attention to key cycles and key orders
on symmetric keys. Moreover, we consider atomic keys for symmetric encryption since
there exists no general definition (with a cryptographic interpretation) of key cycles in the case of arbitrary composed keys and soundness results are usually obtained for atomic keys.


More precisely, we assume a sort
$\keys\subset\term$ and we assume that the sort of $\enc$ is
$\term\times \keys \rightarrow \term$. All the other symbols are
of sort $\term\times \cdots \times \term \rightarrow \term$. Hence only names and variables can be of
sort $\keys$. In this section we call \emph{key} a variable or a name of sort $\keys$. Finally, for any list of terms $L$, $\lset{L}$ is the set of terms
that are members of the list.
In this section, we consider (in)security properties of the form
$P(L)$ where $P$ is a predicate symbol and $L$ is a list of terms. Informally,
$\sigma$ will be a solution of $P(L)$ if $\lset{L}\sigma$ contains a key cycle.
The precise interpretation of $P$ depends on the notion of key-cycle:
this is what we investigate first in the following section.

\subsection{Key cycles}
Many definitions of key cycles are available in the literature. They
are stated in terms of an ``encryption'' relation between keys or
occurrences of keys.  
An early definition proposed by Abadi and Rogaway~\cite{ARCryptology02},
identifies a key cycle with a cycle in the encryption relation, with
no conditions on the occurrences of the keys.
However, the definition induced by Laud's
approach~\cite{Laud-NORDSEC02} corresponds to searching for such
cycles only in the ``visible'' parts of a message. For example the
message $\penc{\penc{k}{k}}{k'}$ contains a key cycle using the former
definition but does not when using the latter one and assuming that
$k'$ is secret. It is generally admitted that the Abadi-Rogaway
definition is unnecessarily restrictive and hence we will say that the
corresponding key cycles are \emph{strict}. However, for completeness
reasons, we treat both cases.



There can still be other variants of the definition,
depending on whether  the relation ``$k$ encrypts $k'$'' is restricted or not to keys $k'$ that occur in
plain-text. For example, $\enc(\enc(a,k),k)$ may or may not contain a key cycle. As above, even if
occurrences of keys used for encrypting (as $k$ in $\penc{m}{k}$) need not be considered as encrypted keys,
and hence can safely be ignored when defining key cycles, we consider both cases. Note that the
initial Abadi-Rogaway setting considers that $\enc(\enc(a,k),k)$ has a key cycle.

We write $s<_{st} t$ if and only if $s$ is a subterm of $t$.
$\sqsubseteq$ is the least reflexive and transitive relation
satisfying: $s_1\sqsubseteq
(s_1,s_2)$, $s_2\sqsubseteq (s_1,s_2)$, and, if $s\sqsubseteq t$,
 then $s\sqsubseteq \penc{t}{t'}$.
Intuitively, $s\sqsubseteq t$ if $s$ is a subterm of $t$ that either
occurs (at least once) in clear (i.e.~not encrypted) or occurs (at
least once) in a plain-text position.  A position $p$ is a
\emph{plain-text position} in a term $u$ if there exists an occurrence
$q$ of an encryption in $u$ such that $q\cdot 1\le p$.

\begin{definition}
Let $\rhoun$ be a relation 
chosen in
$\{<_{st},\sqsubseteq\}$. Let $S$ be a set of terms and $k, k'$ be two keys. We say that
$k$ \emph{encrypts} $k'$ \emph{in} $S$ (denoted $k\, \rho_e^S\, k'$) if there exist
$m\in S$ and a term $m'$ such that $$k'\rhoun m'\mbox{ and } \enc(m',k)\sqsubseteq m.$$
\end{definition}
For simplicity, we may write $\rho_e$ instead of $\rho_e^S$, if $S$ is
clear from the context.  Also, if $m$ is a message we denote by
$\rho_e^m$ the relation $\rho_e^{\{m\}}$.


Let $S$ be a set of terms. We define $\hidden{S}\!\eqdef\!\set{k\in\st(S)\mid k\mbox{ of sort }\keys,
S\not\vdash k}$.

\begin{definition}[(Strict key cycle)]\label{def:skc}
Let $\HK$ be a set of keys. We say that a set of terms $S$ contains a \emph{strict key
cycle} on $K$ if there is a cycle in the restriction of the relation $\rho_e^S$ on $K$. Otherwise we say that
$S$ is \emph{\sacyclic} on $K$.

We define the predicate $P_{skc}$ as follows:
$L\in P_{skc}$ if and only if
the set $\set{m\mid \lset{L}\vdash m}$ contains a strict key cycle on $\hidden{\lset{L}}$.
\end{definition}


We give now the definition induced by Laud's approach~\cite{Laud-NORDSEC02}. He has showed in a passive
setting that if a protocol is secure when the intruder's power is given by a modified Dolev-Yao deduction
system $\vdash_{\emptyset}$, then the protocol is secure in the computational model, without requiring a ``no
key cycle'' condition. Rephrasing Laud's result in terms of the standard deduction system $\vdash$ gives rise
to the definition of key cycles below, as it has been proved in~\cite{Janvier-these}.

To state the following definition we need a more precise notion than the encrypts relation. We say that
an occurrence $q$ of a key $k$ \emph{is protected} by 
a key $k'$ in a term $m$
if $m|_{q'}=\penc{m'}{k'}$ for some term $m'$ and some position $q'$, and the occurrence of $k$ at $q$ in
$m$ is a plain-text
occurrence of $k$ in $m'$, that is $q'\cdot 1\le q$. 
We extend this definition in the intuitive way to sets of terms. This can be done for example
by indexing the terms in the set and adding this index as a prefix to the position in the term to
obtain the position in the set.

\begin{definition}[(Key cycle~\cite{Janvier-these})]\label{def:kc}
Let $K$ be a set of keys. We say that a set of terms $S$ is \emph{acyclic} on $K$ if there
exists a strict partial ordering
$\sko$ on $K$ such that for all $k\in K$, for all occurrences $q$ of $k$ in plain-text position in $S$, there
is $k'\in K$ 
such that $k'\sko k$ and $q$ is protected by $k'$ in $S$. Otherwise we say that $S$ contains a \emph{key cycle} on $K$.

We define the predicate $P_{kc}$ as follows: for any list of terms $L$,
$L\in P_{kc}$ if and only if
the set $\set{m\mid \lset{L}\vdash m}$ contains a key cycle on $\hidden{\lset{L}}$.
\end{definition}

We say that a term $m$ contains a (strict) key cycle if the set $\{m\}$ contains one.

\begin{example}
The messages $m=\enc(\enc({k},{k}),{k'})$ and
$m'\!=\!\langle{\enc(k_1,k_2)},$ $\enc(\enc(k_2,$ $k_3),k_1)\rangle$
are acyclic, while the message
$m''\!=\pair{\pair{\enc(k_1,k_2)}{\enc(\enc(k_2,k_1),k_3)}}{k_3}$ has
a key cycle. The orderings $k'\sko k$ and $k_3\sko k_2\sko k_1$ prove it
for $m$ and $m'$ while for $m''$ such an ordering cannot be found since
$k_3$ is deducible. However, all three messages have strict key
cycles.
\end{example}

\subsection{Key orderings}
In order to establish soundness of formal models in a symmetric
encryption setting, the requirements on the encrypts relation
can be even stronger, in particular in the case of an active
intruder. In~\cite{Backes_Pfitzmann_CSFW04_symmetric_encryption}
and~\cite{cryptoeprint:2005:020} 
the authors require that a key never
encrypts a younger key. More precisely, the encrypts relation has to
be compatible with the ordering in which the keys are generated. Hence we also want to check whether there
exist executions of the protocol for which the encrypts relation is incompatible with an \textit{a priori}
given order on keys.

\begin{definition}[(Key ordering)]\label{def:korder}
Let $\sko$ be a strict partial ordering on a set of  keys $K$. We say that a set of terms $S$
is \emph{compatible} with $\sko$ on $K$ if
\[ k\,\rho_e^{S}\, k'\ \Rightarrow\ k'\not\nko k, \text{ for all }k,k'\in\HK.\]

Given a strict partial ordering $\sko$ on a set of keys, we define the predicate $P_{\sko}$ as follows:
$P_{\sko}$ holds on a list of terms $L$ if and only if
the set $\set{m\mid \lset{L}\vdash m}$ is compatible with $\sko$ on $\hidden{\lset{L}}$.
\end{definition}
For example,
in~\cite{Backes_Pfitzmann_CSFW04_symmetric_encryption,cryptoeprint:2005:020}
the authors choose $\sko$ to be the order in which the keys are
generated: $k\sko k'$ if $k$ has been generated before $k'$.
We denote by $\overline{P}_{\sko}$  the
negation of $P_{\sko}$. 
Indeed, an attack in this context is an execution such that the
encrypts relation is incompatible with~$\sko$. 


\subsection{Properties that are independent of the notion of key cycle}
We show how to decide the existence of key cycles or the conformation
to an ordering in polynomial time for solved \dedsyss{}.
Note that the set of messages on which our predicates are applied usually contains all messages sent on
the network and possibly some additional intruder knowledge.

We start with statements, that do not depend on which notion of key cycle
we choose.

\begin{lemma}\label{lem:protects_deduc2}
Let $S$ be a set of terms, $m$ be a term and $k$ be 
a key such that $S\vdash m$ and~$S\not\vdash k$. Then for
any plain-text occurrence $q$ of $k$ in $m$, there is a plain-text occurrence $q_0$ in $S$ such that,
if there is key $k'$ with $S\not\vdash k'$, and
which protects $q_0$ in $S$, then $k'$ protects~$q$~in~$m$.
\end{lemma}
\begin{proof}
We reason by induction on the depth of the proof of $S\vdash m$:
\begin{itemize}
 \item if the last rule is an axiom, then  $m\in S$. We may simply
choose $q_0=q$.
 \item if the last rule is a decryption, 
then $S\vdash \penc{m}{k''}$ and $S\vdash k''$ for some $k''\neq k$.
Take the position $q_1=1\cdot q$ in $\penc{m}{k''}$. It is an occurrence of $k$. Applying the induction
hypothesis we obtain an occurrence $q_0$ of $k$ in $S$ such that, if there is a key
$k'$ with $S\not\vdash k'$ and
which protects $q_0$ in $S$, then
$k'$ protects $q_1$ in  $\penc{m}{k''}$. Since $S\not\vdash k'$, it
follows that $k''\neq k'$ and hence $k'$ protects $q$ in $m$.
 \item if the last rule is a another rule, we  proceed in a similar
way as above.
\end{itemize}
\end{proof}

As a corollary we obtain the following proposition, which
 states that, in the passive case, a key cycle can be
deduced from a set $S$ only if it already appears in $S$.
\begin{proposition}\label{prop:prop_deduc}
Let $L$ be a list of ground terms, and $\sko$ a strict partial ordering on a set of keys. The
predicate  $P_{kc}$ (respectively, $P_{skc}$ or $\overline{P}_{\sko}$) holds on
$L$ if and only if $\lset{L}$ contains a key cycle (respectively, $\lset{L}$ contains a strict key cycle, or
the encrypts relation on $\lset{L}$ is not compatible with~$\sko$). 
\end{proposition}
\begin{proof}
The right to left direction is trivial since $\lset{L}\subseteq\{m\mid\lset{L}\vdash m\}$.

We will prove the left to right direction only for the key cycle property,
 the other two properties can be proved in a similar way. 
Assume that there is no strict partial ordering satisfying the conditions in
Definition~\ref{def:kc} for $\{m\mid\lset{L}\vdash m\}$. In other words, for any strict partial ordering $\sko$
on $\hidden{\lset{L}}$ there is a key $k$ and an occurrence $q$ of $k$ in $\{m\mid\lset{L}\vdash m\}$ such
that for any key $k'$,  $k'$ protects $q$ in $\{m\mid\lset{L}\vdash m\}$ implies $k'\not\sko k$. Using the
previous lemma we can replace $\{m\mid\lset{L}\vdash m\}$ by $\lset{L}$ in the previous sentence, thus
obtaining that there is a key cycle in $\lset{L}$.
\end{proof}



\comment{Let  $C$ be a solved constraint system, $theta$ be a {\simple} solution of $C$ and $L$ be a list of
messages
such that $\var(\lset{L})\subseteq\var(C)$ and $\lleft{C}\subseteq \lset{L}$. Also let $\HK$ be a set of
names of sort $\keys$.  Let $\le$ be a partial ordering on $\HK$. If there is $k\in\HK$ such that $k$ is
deducible, that is $\lleft{C}\theta\vdash k$ then the predicates $P_{kc}^{\HK}$ and
$\overline{P}_{\le}({\HK})$ hold. If all the keys in $\HK$ are not deducible then $P_{kc}^{\HK}$ holds on
$L\theta$ if and only if
}

The next lemma will be used to show that $\hidden{\lset{L}\theta}$ does not
depend on the 
solution $\theta$ of a solved constraint $C$.
\begin{lemma}\label{lem:similar_deduc}
Let $T\Vdash x$ be a constraint of a solved constraint system $C$,
$\theta$ a  solution of $C$ and $m$ a non-variable term. If $T\theta\vdash m$ then there is a
non-variable term $u$ with $\var(u)\subseteq\var(T)$ such that $T\cup{\var(T)}\vdash u$ and $m=u\theta$.
\end{lemma}
\begin{proof}
We write $C$ as $\bigwedge_i(T_i\Vdash x_i)$, with $1\le i\le n$ and $T_i\subseteq T_{i+1}$.
Consider the index $i$ of the constraint $T\Vdash x$, that is such that $(T_i\Vdash u_i)\in C$, $T_i=T$ and $u_i=x$. The lemma is proved by induction on $(i,l)$ (considering the lexicographical ordering)
where $l$ is the length of the proof of $T_i\theta\vdash m$. Consider the last rule of the proof:
\begin{itemize}
 \item (axiom rule) $m\in T_i\theta$. Then there is $u\in T_i$ such that $m=u\theta$. If $u$ is a variable
then there is $j<i$ such that $T_j\Vdash u$ is a constraint of $C$. We have $T_j\theta\vdash u\theta$. Then
by induction hypothesis there is a non-variable term $u'$ with $\var(u')\subseteq\var(T_j)$ such that
$T_j\cup\var(T_j)\vdash u'$ and $u\theta=u'\theta$. Hence $u'$ satisfies the conditions.

 \item (decomposition rule) Suppose the rule is the decryption rule. Then the premises of the rule are
$T_i\theta\vdash \penc{m}{k}$ and $T_i\theta\vdash k$ for some term $k$. By induction hypothesis
there are non-variable terms $u_1$ and $u_2$ with $\var(u_1),\var(u_2)\subseteq\var(T_i)$ such that
$T_i\cup{\var(T_i)}\vdash u_1$, $T_i\cup{\var(T_i)}\vdash u_2$, $u_1\theta=\penc{m}{k}$ and
$u_2\theta=k$. Then $u_1=\penc{u}{u'_2}$ with $u\theta=m$ and $u'_2\theta=k$. If $u$ is a variable then, as in
the previous case, we find an $u'$ satisfying the conditions. Suppose $u$ is not a variable. We still need to
show that $T_i\cup{\var(T_i)}\vdash u$. If $u'_2$ is a variable then
$T_i\cup{\var(T_i)}\vdash u'_2$ since $u'_2\in\var(T_i)$. If $u'_2$ is not a variable then $u'_2\theta=u'_2$
hence $u'_2=u_2$. In both cases it follows that $T_i\cup{\var(T_i)}\vdash u$. The projection rule case is
simpler and is treated similarly.

\item (composition rule) This case follows easily from the induction hypothesis applied on the premises.
\end{itemize}
\end{proof}


\begin{corollary}\label{cor:same_keys}
Let $T\Vdash x$ be a constraint of a solved \dedsys{} $C$, and $\theta$, $\theta'$ be two 
solutions of $C$. Then for any key $k$, $T\theta\vdash k$ if and only if $T\theta'\vdash k$.
\end{corollary}
\begin{proof}
Suppose that $T\theta\vdash k$. From the previous lemma we obtain that there is a non-variable $u$ with
$\var(u)\subseteq\var(T)$ such that $T\cup{\var(T)}\vdash u$ and $k=u\theta$.
Since keys are atomic and $\theta$ is a ground substitution it follows that $u=k$. Hence
$T\theta'\cup\set{x\theta'\mid x\in\var(T)}\vdash k$. So $T\theta'\vdash k$, since $\theta'$ is a  solution (and thus $T\theta'\vdash x\theta'$ for all $x\in\var(T)$) and by using Lemma~\ref{lemma_deduc-cutelim}.
\end{proof}

\subsection{Decision results}

On solved \dedsyss{}, it is possible to decide in polynomial time,
whether an attacker can trigger a key cycle or not, whatever notion of
key cycle we consider: 

\begin{proposition}\label{lemma_deter_ext}
Let  $C$ be a solved \dedsys{}, $L$ be a list of messages such that
$\var(\lset{L})\subseteq\var(C)$ and $\lleft{C}\subseteq \lset{L}$, and $\sko$ a strict partial ordering on a
set of keys. Deciding whether there exists an attack for $C$ and $P(L)$  can be done in $\mathcal{O}(|L|^2)$, for any $P\in \{P_{kc}, P_{skc},\overline{P}_{\sko}\}$.
\end{proposition}



We devote the remaining of this section to the proof of the above proposition.

We know by Proposition~\ref{prop:prop_deduc} that it is sufficient to analyze the encrypts (or protects)
relation only on $\lset{L}\theta$ (and not on every deducible term), where $\theta$ is an arbitrary
 solution.

We can safely assume that there is exactly one \dedcons{} for each variable. 
Indeed, eliminating from $C$ all constraints $T'\Vdash x$ 
for which there is a constraint $T\Vdash x$ in $C$ with $T\subsetneq T'$ 
we obtain an equivalent \dedsys{} $C'$ : $\sigma$ is a solution of $C'$ iff it is a solution of $C$.
Let $t_x$ be the term obtained by pairing all terms of $T_x$ (in some arbitrary ordering).
We write $C$ as $\bigwedge_i(T_i\Vdash x_i)$, with $1\le i\le n$ and $T_i\subseteq T_{i+1}$.
We construct the following substitution $\tau=\tau_1\dots\tau_n$,
and $\tau_j$ is defined inductively as follows:
\begin{itemize}
\item[-] $\dom(\tau_1)=\set{x_1}$ and $x_1\tau_1=t_{x_1}$
\item[-] $\tau_{i+1}=\tau_i\cup\set{\subst{x_{i+1}}{t_{x_{i+1}}\tau_i}}$.
\end{itemize}
The construction is correct by the definition of \dedsyss{}.
It is clear that $\tau$ is a  solution of $C$. We show next that it is sufficient to analyze this
particular solution. 


\subsubsection{Key cycles} 
We focus first on the property $P_{kc}$.

\begin{lemma}\label{lem:tau_sol_kc}
Let $C$ be a solved \dedsys{}, $L$ a list of terms such that $\var(L)\subseteq\var(C)$,
$\lhs(C)\subseteq\lset{L}$, and assume $P$ is interpreted as $P_{kc}$. Then
there is an attack for $C$ and $P(L)$ if and only if $\tau$ is an attack
for $C$ and $P(L)$.
\end{lemma}
\begin{proof}
We have to prove that if there is no partial ordering satisfying the conditions in Definition~\ref{def:kc} for
the set $\lset{L}\theta$ (according to Proposition~\ref{prop:prop_deduc}) then there is no partial ordering
satisfying the same conditions for $\lset{L}\tau$. Suppose that there is a strict partial ordering $\sko$ which
satisfies the conditions for $\lset{L}\tau$. We prove that the same partial ordering does the job for
$\lset{L}\theta$.

Let $C'=C\wedge(\lset{L}\Vdash z)$ where $z$ is a new variable. $C'$ is a 
\dedsys{} since
$\lhs(C)\subseteq\lset{L}$.
We write $C'$ as $\bigwedge_i(T_i\Vdash x_i)$, with $1\le i\le n$ and $T_i\subseteq T_{i+1}$.
We prove by induction on $i$ that for all $k\in\hidden{\lset{L}\theta}$, for all plain-text occurrences $q$ of $k$ in $T_i\theta$ there is a key
$k'\in\hidden{\lset{L}\theta}$ 
such that $k'\sko k$ and $k'$ protects $q$ in $T_i\theta$. It is sufficient to prove this since for $i=n$ we have $T_i=\lset{L}$.
Remark also that from Corollary~\ref{cor:same_keys} applied to $\lset{L}\Vdash z$ we obtain that
$\hidden{\lset{L}\theta}=\hidden{\lset{L}\tau}$.

For $i=1$ we have $T_1=T_1\theta=T_1\tau$ hence the property is clearly satisfied for $\theta$ since it is
satisfied for $\tau$.

Let $i>1$. Consider an occurrence $q$ of a key $k\in\hidden{\lset{L}\theta}$ in a plain-text position of $w$
for some $w\in T_i\theta$. Let $t\in T_i$ such that $w=t\theta$.

If $q$ is a non-variable position in $t$ then it is a position in $t\tau$.
And since $\tau$ is a solution we have that there is 
a key $k'\in\hidden{\lset{L}\tau}$ (hence $k'\in\hidden{\lset{L}\theta}$)
such that $k'\sko k$ and $q$ is protected by $k'$ in $t\tau$. The key $k'$ cannot occur  in some $x\tau$, with
$x\in\var(t)$, since otherwise $k'$ is deducible (indeed $x\tau=k'$ since the keys are atomic and
$T_x\tau\vdash x\tau$). Hence $k'$ occurs in $t$. Then $k'$ protects $q$ in $t$, and thus in $w$ also.

If $q$ is not a non-variable position in $t$ then there is a variable $x_j\in\var(t)$ with $j<i$ such that
the occurrence $q$ in $t\theta$ is an occurrence of $k$ in $x_j\theta$
(formally $q=p\cdot q'$ where $p$ is some position of $x_j$ in $t$ and $q'$ is some occurrence of $k$ in $x_j\theta$).
Applying Lemma~\ref{lem:protects_deduc2} we obtain that there is an occurrence $q_0$ of $k$ in
$T_j\theta$ such that if there is a key $k'$ with $T_j\theta\not\vdash k'$ and
which protects $q_0$ in $T_j\theta$ then 
$k'$ protects $q'$ in $x_j\theta$. The existence of the
key $k'$ 
is assured by the induction hypothesis on $T_j\theta$. Hence $k'$ protects $q'$ in $x_j\theta$ 
and thus $q$ in $w$.
%
since otherwise there is $x\in\var(\lset{L})$ such that $x\tau=k'$, which implies that
$k'\notin\hidden{\lset{L}}$. Then $q'$ is a position in $\lset{L}\theta$. Moreover $q'$ protects $q$ in
$\lset{L}\theta$.

If $q$ is not a non-variable position in $\lset{L}$ then there is a variable $x\in\var(\lset{L})$ such that
\end{proof}


Hence we only need to check whether $\tau$ is an attack for $C$ and $P(L)$. Let 
$K=\hidden{\lset{L}\tau}$. We build inductively the sets $K_0=\emptyset$ and for all
$i\ge 1$,
$$K_{i}=\set{k\in K\mid \forall q\in\posp(k,\lset{L}\tau)\, \exists k'\ \mbox{s.t.~$k'$ protects $q$
and } k'\in K_{i-1}}$$
where $\posp(m,T)$ denotes the plain-text positions of a term $m$ in a set $T$.
Observe that for all $i\ge 0$, 
$K_i\subseteq K_{i+1}$. This can be proved easily by induction on $i$. Moreover, since $K$ is finite and $K_i\subseteq K$ for all $i\ge 0$,
then there is $l\ge 0$ such that $K_i=K_l$ for all $i> l$.

\begin{lemma}
There exists $i\ge 0$ such that $K_i=K$ if and only if $L\tau\in P_{kc}$.
\end{lemma}
\begin{proof}
Consider first that there exists $i\ge 0$ such that $K_i=K$. Then take the following strict partial ordering on
$K$: $k'\sko k$ if and only if there is $j\ge 0$ such that $k'\in K_j$ and $k\notin K_j$. Consider a key
$k\in K$ and a plain-text occurrence $q$ of $k$ in $\lset{L}\tau$. Then take $l\ge 1$ minimal such that
$k\in K_l$. By the definition of $K_l$ there is $k'\in K$ such that $k'$ protects $q$ and $k'\in K_{l-1}$.
Since $l$ is minimal $k\notin K_{i-1}$. Hence $k'\sko k$. Thus $L\tau\in P_{kc}$. 

Consider now that $\tau$ is a solution. 
Suppose that $K_{i+1}=K_i\subsetneq K$. Let $k\in K\setminus K_{i+1}$. Since $k\not\in K_{i+1}$ there is a plain-text occurrence $q$
of $k$ such that for all $k'\in K$ either $k'$ does not protect $q$, or $k'\notin K_{i}$. But
since $\tau$ is a solution, there is $k''\in K$ such that $k''$ protects $q$ and $k''\sko k$. It follows that
$k''\notin K_{i}$, and thus $k''\notin K_{i+1}$. Hence for an arbitrary $k\in K\setminus K_{i+1}$ we have
found $k''\in K\setminus K_{i+1}$ such that $k''\sko k$. That is, we can build an infinite sequence
$\dots \sko k'' \sko k$ with distinct elements from a finite set -- contradiction. So there exists $i\ge 0$
such that $K_i=K$.
\end{proof}

Hence to check whether $L\tau\in P_{kc}$, we only need to construct the sets $K_i$ until
$K_{i+1}=K_{i}$ and then to check whether $K_i=K$. This algorithm is similar to a classical method for
finding a topological sorting of vertices (and for finding cycles) of directed graphs. 
It is also similar to
that given by Janvier~\cite{Janvier-these} for the intruder deduction problem considering the deduction
system of Laud~\cite{Laud-NORDSEC02}.

Regarding the complexity, there are at most $\card{K}$ sets to be build and each set $K_i$ can be constructed in
$\mathcal{O}(|\lset{L}\tau|)$. If a DAG-representation of the terms is used then
$|\lset{L}\tau|\in\mathcal{O}(|\lset{L}|)$. This gives a complexity of $\mathcal{O}(|K|\times |\lset{L}|)$
for the above algorithm.

\medskip

\subsubsection{Strict key cycles and key orderings.}
For the other two properties $P_{skc}$ and $\overline{P}_{\sko}$ we proceed in a similar manner.

\begin{lemma}\label{lem:encrypts_deduc}
Let $T\Vdash x$ be a constraint of a solved \dedsys{} $C$ and $\theta$ be a
 solution. Let $m, u, k$ be terms such that
\[
T\theta\vdash m \mbox{ and } \enc(u,k)\sqsubseteq m \mbox{ and }
T\theta\not\vdash k. 
\]
Then there exists a non-variable term $v$ such that  $v\sqsubseteq w$ for some $w\in T$
and $v\theta = \enc(u,k)$.
\end{lemma}
\begin{proof}
We write $C$ as $\bigwedge_i(T_i\Vdash x_i)$, with $1\le i\le n$ and $T_i\subseteq T_{i+1}$.
Consider the index $i$ of the constraint $T\Vdash x$, that is such that $T_i\Vdash u_i\in C$, $T_i=T$ and
$u_i=x$. The lemma is proved by induction on $(i,l)$ (lexicographical ordering)
where $l$ is the length of the proof of $T_i\theta\vdash m$.
Consider the last rule of the proof:
\begin{itemize}
\item (axiom rule) $m = t\theta$ for some $t\in T_i$. We can have that either there is $t'\sqsubseteq t$ such
that $t'\theta=\enc(u,k)$, or $\enc(u,k)\sqsubseteq y\theta$ for some $y\in\var(t)$. In the first case
take $v=t'$, $w=t$. In the second case, by the definition of \dedsyss{},
there exists $(T_j\Vdash y)\in C$ with $j<i$. Since $T_j\theta\vdash
y\theta$ and $T_j\theta\not\vdash k$ (since $T_j\subseteq T_i$), we
deduce by induction hypothesis that there exists a non-variable term
$v$ such that $v\sqsubseteq w$ for some $w\in T_j$, hence $w\in T_i$ and $v\theta = \enc(u,k)$.
\item (decomposition rule) Let $m'$ be the premise of the rule. We have that $T_i\theta\vdash m'$ (with a
  proof of a strictly smaller length) and $m\sqsubseteq m'$
thus $\enc(u,k)\sqsubseteq m'$. By induction hypothesis, we deduce that there
exists a non-variable term $v$ such that  $v\sqsubseteq w$ for some $w\in T_i$ and $v\theta
= \enc(u,k)$.
\item (composition rule) All cases are similar to the previous one
  except if $m= \enc(u,k)$ and the rule is {\small
$\displaystyle\frac{S\vdash x\quad S\vdash y}{S\vdash \enc(x,y)}$}.
But this case contradicts $T_i\theta\not\vdash k$.
\end{itemize}
\end{proof}


The following simple lemma is also needed for the proof of Lemma~\ref{lem:tau_sol_encrypts}.
\begin{lemma}\label{lem:rho}
Let $T\Vdash x$ be a constraint of a solved \dedsys{} $C$, $\theta$ be a
solution, $k\in\hidden{T\theta}$, and $m$ a term such that $T\theta\vdash m$.
If $k\rhoun m$ then there is $t\in T$ such that $k\rhoun t$.
\end{lemma}
\begin{proof}
We write $C$ as $\bigwedge_i(T_i\Vdash x_i)$, with $1\le i\le n$ and $T_i\subseteq T_{i+1}$.
Consider the index $i$ of the constraint $T\Vdash x$, that is such that $(T_i\Vdash u_i)\in C$, $T_i=T$ and $u_i=x$. The lemma is proved by induction on $(i,l)$ (considering the lexicographical ordering)
where $l$ is the length of the proof of $T_i\theta\vdash m$. Consider the last rule of the proof:
\begin{itemize}
 \item (axiom rule) $m\in T_i\theta$ or $m$ a public constant. If $m$ is a public constant then $k\neq m$ since $k\in\hidden{T\theta}$. Thus there is $t\in T_i$ such that $m=t\theta$. If $k\rhoun t$ then we're done. Otherwise there is a variable $y\in\var(t)$ such that $k\rhoun y\theta$. Also, there is $j<i$ such that $T_j\Vdash y$ is a constraint of $C$. Then, by induction hypothesis, there is $t'\in T_j$, hence in $T_i$, such that $k\rhoun t'$.

 \item (composition or decomposition rule) By inspection of all the composition and decomposition rules we observe that there is always a premise $T_i\theta\vdash m'$ with $k\rhoun m'$ for some term $m'$. The conclusion follows then directly from the induction hypothesis.
\end{itemize}
\end{proof}

The
following lemma shows that it is sufficient to analyze $\tau$ when checking the properties $P_{skc}$ and
$\overline{P}_{\sko}$.
\begin{lemma}\label{lem:tau_sol_encrypts}
Let $C$ be a solved \dedsys{}, $L$ a list of terms such that $\var(L)\subseteq\var(C)$ and
$\lhs(C)\subseteq\lset{L}$, and $\theta$ a solution of $C$. For any
$k,k'\in\hidden{\lset{L}\theta}$, if $k$ encrypts $k'$ in $\lset{L}\theta$ then $k$ encrypts
$k'$ in $\lset{L}\tau$.
\end{lemma}
\begin{proof}
Remember that $\hidden{\lset{L}\theta}=\hidden{\lset{L}\tau}$ (Corollary~\ref{cor:same_keys}).

Consider two keys $k,k'\in\hidden{\lset{L}\theta}$ such that $k$ encrypts $k'$ in $\lset{L}\theta$.
Then there are terms $u, u'$ such that $u'\in \lset{L}\theta$, $\penc{u}{k}\sqsubseteq u'$ and
$k'\rhoun u$. We can have that either (first case) there are $v, w$ such that $v\sqsubseteq w\in \lset{L}$,
$v$ non-variable and $\enc(u,k)=v\theta$, or (second case) $\enc(u,k)\sqsubseteq x\theta$ with
$x\in\var(\lset{L})$. In the second case, consider the constraint $(T_x\Vdash x)\in C$. We have
$T_x\theta\vdash x\theta$. Hence we can apply Lemma~\ref{lem:encrypts_deduc} for $x\theta$, $u$
and $k$ to obtain that there exists a non-variable term $v$ such that $v \sqsubseteq w$ for some $w\in T_x$
and $v\theta = \enc(u,k)$. Hence, in both cases, we obtained that there is a non-variable term
$v\in\st(\lset{L})$ (since $T_x\subseteq \lset{L}$) such that $v\theta=\penc{u}{k}$. Thus there is $v_0$
such that $v=\penc{v_0}{k}$. Indeed, otherwise $v=\penc{v_0}{y}$ for some $y\in\var(\lset{L})$, hence $y\in\var(C)$. Since $C$ is solved we have $T_y\sigma\vdash y\sigma$. But $y\sigma=k$, contradicting $k\in\hidden{\lset{L}\theta}$.

We have $v_0\theta=u$. Since $k'\rhoun u$ and $k'$ is a name or a variable, we can have that
$k'\rhoun v_0$, or $k'\rhoun y\theta$ for some $y\in\var(v_0)$.
If $k'\rhoun v_0$ then $k$ encrypts $k'$ in $\lset{L}$, hence in $\lset{L}\tau$ also.
If $k'\rhoun y\theta$ then from the previous lemma $k'\rhoun t$ for some $t\in T_y$,
and hence $k'\rhoun y\tau$. Therefore in both
cases we have that $k$ encrypts $k'$ in $\lset{L}\tau$.
\end{proof}

We deduce that deciding whether there is an attack for  $C$ and $P(L)$,
when $P$ is interpreted as $P_{skc}$,
can be done simply by
deciding whether the restriction of the relation $\rho_e^{\lset{L}\tau}$ to $K\times K$ is
cyclic.

Deciding whether there is an attack for $C$ and  $P(L)$, when $P$ is interpreted as  $\overline{P}_{\sko}$, can be done by deciding
whether the  restriction to $K\times K$ of the relation $\rho_e^{\lset{L}\tau}$ has the following
property $Q$: there are $k,k'\in K$ such that $k \rho_e^{\lset{L}\tau} k'$ and $k\nko k'$.

Checking the cyclicity of the relation $\rho_e^{\lset{L}\tau}$ reduces to  checking the
cyclicity of the
corresponding directed graph, using a classic algorithm in $\mathcal{O}(|\HK|^2)$. Then,
checking the property
$Q$ can be performed by analyzing all pairs $(k,k')\in K\times K$ hence also in $\mathcal{O}(|\HK|^2)$.

\smallskip



Verifying any of the three properties requires a preliminary step of computing
$K=\hidden{\lset{L}\tau}$. Computing deducible subterms can be performed in linear
time, hence this computation step requires $\mathcal{O}(|\lset{L}\tau|)$. $|\lset{L}\tau|\leq 
|\lset{L}| + |\tau| \leq |\lset{L}| + \mathcal{O}(|C|)$. If $\lhs(C)\subseteq \lset{L}$,
then $|\lset{L}\tau| = \mathcal{O}(|L|)$. It follows that the complexity of deciding
whether there is an attack for $C$ and $P(L)$ is $\mathcal{O}(|L|^2)$, when $P$ is
interpreted as $P_{kc}, P_{skc}$ or $\overline{P}_{\sko}$.

\subsection{NP-completeness} Let $C$ be a \dedsys{} and $L$ a list of terms such that
$\var(\lset{L})\subseteq\var(C)$ and $\lhs(C)\subseteq\lset{L}$.
The NP membership of deciding whether there is an attack for $C$ and $P(L)$
(for our 3 possible interpretations of $P$) 
follows immediately from Corollary~\ref{corol:general}
and Proposition~\ref{lemma_deter_ext}.

NP-hardness is obtained by adapting the construction for NP-hardness provided
in~\cite{RT03TCS}. 
More precisely, we consider the reduction of the 3SAT problem to our problem. For any 3SAT Boolean formula
we construct a protocol such that the intruder can deduce a key cycle if and only if the formula is
satisfiable. The construction is the same as in~\cite{RT03TCS} (pages 15 and 16) except that, in the last
rule, the participant responds with the term $\enc(k,k)$, for some fresh key $k$ (initially secret), instead
of $Secret$. Then it is easy to see that the only way to produce a key cycle on a secret key is to play this
last rule which is equivalent, using~\cite{RT03TCS}, to the satisfiability of the corresponding 3SAT
formula.



\section{Authentication-like properties}\label{sec:auth}
We propose a simple decidable logic for security properties. This
logic enables in particular to specify authentication-like properties.

\subsection{A simple logic}
The logic enables terms comparisons and is closed under Boolean connectives.


\begin{definition}
The logic $\logic$ is inductively defined by:
\[\phi ::= [m_1=m_2] \mid \neg \phi \mid \phi\vee\phi \mid
\phi\wedge\phi \mid \bot
\quad\quad\quad m_1,m_2 \mbox{ terms}
\]
$\var(\phi)$ is the set of variables occurring in its atomic formulas.
\end{definition}

$\sigma \models [m_1=m_2]$ if $m_1\sigma$ and $m_2\sigma$ are identical terms.
$\sigma\not\models \perp$.
This satisfaction relation is extended to any of the above formulas, 
interpreting the Boolean connectives as usual.




\begin{example}
Let us consider again the authentication property introduced in
Example~\ref{ex:auth}. There is an attack on authentication between
$A$ and $B$  if $A$ and $B$ do not agree on the nonce $n_a'$ sent by
$A$ for $B$, that is if $x=n_a'$
at the end of the run of the protocol.
This can be expressed by the following formula
\[\phi_1 = [x\neq n_a']\]
The substitution $\sigma_1$ (assigning $x$ to $n_a$) is an
attack for $C_1'$ (defined in Example~\ref{ex:auth}) and $\phi_1$ and demonstrates
a failure of authentication.
\end{example}

More sophisticated properties can be expressed using the logic
$\logic$. For example, when two sessions of the same role are executed, one can
expressed that an agent has received \emph{exactly once} the right
nonce $n_a$, with the following formula.
\[\phi_2 = ([x_1=n_a]\wedge[x_2\neq n_a])\vee([x_1\neq n_a]\wedge[x_2= n_a])\]
where $x_1$ (resp. $x_2$) represents the nonce received by the agent in the
first (resp. second) session.

We can also express properties of the form: if two agents agree on
some term $u$, they also agree on some term $v$. This can be indeed
modeled by the formula
\[\phi_3 = [u_1=u_2]\rightarrow [v_1=v_2]
\]
where $u_1$ (resp. $u_2$) represents the view of $u$ by the first
(resp. second) agent and
$v_1$ (resp. $v_2$) represents the view of $v$ by the first
(resp. second) agent. The formula $A\rightarrow B$ is the usual
notation for the formula $\neg A\vee B$.

\subsection{Decidability}

\begin{theorem}
Let $C$ be a \dedsys{} and $\phi$ be a formula of $\logic$.
Deciding whether there is an attack for $C$ and $\phi$ can be performed in
non-deterministic polynomial time.
\end{theorem}
\begin{proof}

First, choosing non-deterministically $\phi_1$ or $\phi_2$ in any
subformula $\phi_1\vee \phi_2$, we may, w.l.o.g.~only consider the
case where $\phi$ is a conjunction $\bigwedge_j [u_j = u'_j] \wedge \phi_d$,
where $\phi_d= \bigwedge_l 
[v_l\neq v'_l]$. 



Let $\sigma$ be a $\mgu$~(idempotent, which does not introduce new variables)
of $\bigwedge_j u_j=u'_j$.  
The \dedsys{} $C$ has a joined solution with $\phi$ if and only if
$C\sigma$ and $\phi_d\sigma$ have a common solution. As in the previous sections,
we choose a representation of expressions, such that applying a $\mgu$~of
subterms of an expression $e$ on $e$ does not increase the size of the expression $e$. 

We are now left to the case where we have to decide whether a
\dedsys{} has a solution together with a property of the form
$\phi = \bigwedge_{i=1}^k [u_i\neq   v_i]$.


Applying Theorem~\ref{theo:general},  there exists
a solution $\theta$ of $C$ and $\phi$ 
if and only if there exist a \dedsys{} $C'$ in solved form
  and  substitutions $\sigma,\theta'$ such that $\theta
  =\sigma\theta'$, $C\simpl^*_{\sigma} C'$ and $\theta'$ is an attack for $C'$ 
and $\phi\sigma$. 
Thus, we are now left to decide whether there exists a solution to
a solved constraint system $C'$ and a formula $\phi\sigma$ of the form
$\phi\sigma = \bigwedge_{i=1}^k [u_i\neq   v_i]$.




If, for some $i$, $u_i$ is identical to $v_i$, then there is clearly no
solution. We claim that, otherwise, there is always a solution. This
is an independence of disequation lemma (as in \cite{colmerauer84} for instance),
and the proof is similar to other independence of disequations lemmas:

\begin{lemma}
Let $C$ be a solved \dedsys{} and $\phi$ be the formula $t_1\neq u_1
\wedge \ldots \wedge t_n\neq u_n$
such that $\var(\phi)\subseteq \var(C)$ and, for every $i$, $t_i$ is not identical to $u_i$. Then  there is always a solution
$\theta$ of $C$ and $\phi$.
\end{lemma}
This is proved by induction on the number of variables of $\phi$.
In the base case, there is no variable and the result is trivial as $\phi$ is
a tautology.

Let $T_0$ be the smallest left-hand
side of $C$. $T_0$ must be a non empty set of ground terms. Note that there is
an infinite set of deducible terms from $T_0$.
%

%
Let $x\in \var(\phi)$. For each $i$, either $t_i=u_i$ has no solution,
in which case $t_i\neq u_i$ is always satisfied, or else let $S= \{
x\sigma_i \mid \sigma_i=\mgu(t_i,u_i)\}$.  We choose $t_x$ such that
$T\vdash t_x$ and $t_x\notin S$. This is possible since $S$ is finite and there are
infinitely many terms deducible from $T$. Now, for every $i$,
$t_i[\subst{x}{t_x}]$ is not identical to $u_i[\subst{x}{t_x}]$ by
construction. Hence, we may apply the induction hypothesis to
$\phi[\subst{x}{t_x}]$ and conclude.
\end{proof}

\section{Timestamps}\label{sec:timestamps}
For modeling timestamps, we introduce a new sort $\Time\subseteq
\term$ for time and we assume an infinite number of names of sort
$\Time$, represented by rational numbers or integers. We assume that the only two sorts are $\Time$ and
$\term$. Any value of
time should be known to an intruder, that is why we add to the
deduction system the rule {\small$\displaystyle\frac{}{S\vdash a}$}
for any name $a$ of sort $\Time$. All the previous results
can be easily extended to such a deduction system since ground
deducibility 
remains decidable in polynomial time.


To express relations between timestamps, we use timed constraints. 
\begin{definition}
An \emph{integer timed constraint} or a
\emph{rational timed constraint} $T$ is a conjunction of formulas of the form
$$\Sigma_{i=1}^k \alpha_i x_i \ltimes \beta,$$
where the $\alpha_i$ and $\beta$ are rational numbers, $\ltimes\in\{<,\leq\}$, and the $x_i$ are variables
of sort $\Time$. A \emph{solution} of a rational (resp. integer) timed constraint $T$ is a closed
substitution $\sigma = \{\subst{x_1}{c_1},\ldots, \subst{x_k}{c_k}\}$, where the $c_i$ are rationals (resp.
integers), that satisfies the constraint.
\end{definition}


Such timed properties can be used for example to say that a
timestamp $x_1$ must be fresher than a timestamp $x_2$ ($x_1\geq
x_2$) or that $x_1$ must be at least 30 seconds fresher than
$x_2$ ($x_1\geq x_2 + 30$).

\begin{example}
We consider the Wide Mouthed Frog Protocol~\cite{CJ97}.
\[
\begin{array}{rl}
A \rightarrow S: & A, \enc(\langle T_a, B, K_{ab}\rangle,K_{as})\\
S \rightarrow B: & \enc(\langle T_s, A, K_{ab}\rangle,K_{bs})
\end{array}
\]
$A$ sends to a server $S$ a fresh key $K_{ab}$ intended for $B$. If the timestamp $T_a$ is fresh enough, the
server answers by forwarding the key to $B$, adding its own timestamps. $B$ simply checks whether this
timestamp is older than any other message he has received from $S$. As explained in~\cite{CJ97}, this
protocol is flawed because an attacker can use the server to keep a session alive as long as he wants by
replaying the answers of the server.

This protocol can be modeled by the following \dedsys{}:
\begin{eqnarray}
S_1\eqdef \{a,b,s,\langle a, \enc(\langle 0, b, k_{ab}\rangle,k_{as})\rangle\}
& \Vdash &
\langle a, \enc(\langle x_{t_1}, b, y_1\rangle,k_{as})\rangle,x_{t_2}\\
S_2\eqdef S_1\cup\{\enc(\langle x_{t_2}, a, y_1\rangle,k_{bs})\}
& \Vdash & \langle b, \enc(\langle x_{t_3}, a, y_2\rangle,k_{bs})\rangle,x_{t_4}\\
S_3\eqdef S_2\cup\{\enc(\langle x_{t_4}, b, y_2\rangle,k_{as})\}
& \Vdash & \langle a, \enc(\langle x_{t_5}, b, y_3\rangle,k_{as})\rangle,x_{t_6}\\
S_4\eqdef S_3\cup\{\enc(\langle x_{t_6}, a, y_3\rangle,k_{bs})\}
& \Vdash & \enc(\langle x_{t_7}, a, k_{ab}\rangle,k_{bs})
\end{eqnarray}
where $y_1,y_2,y_3$ are variables of sort $\term$ and $ x_{t_1},\dots,
x_{t_7}$ are variables of sort $\Time$. We add explicitly the
timestamps emitted by the agents on the right hand side of the
constraints (that is in the messages expected by the participants)
since the intruder can schedule the message transmission whenever he
wants. Note that on the right hand side of constraints we do have
terms, but by abuse of notation we have omitted the pairing function symbol.

Initially, the intruder simply knows the names of the agents and $A$'s message at time~0. Then $S$ answers
alternatively to requests from $A$ and $B$. Since the intruder controls the network, the messages can be
scheduled as slow (or fast) 
as the intruder needs it. The server $S$ should not answer
if $A$'s timestamp is too old (let's say older than 30 seconds) thus $S$'s timestamp cannot be too much
delayed (no more than 30 seconds). This means that we should have $x_{t_2}\leq x_{t_1}+30$. Similarly, we
should have $x_{t_4}\leq x_{t_3}+30$ and $x_{t_6}\leq x_{t_5}+30$. The last rule corresponds to $B$'s
reception. In this scenario, $B$ does not perform any check on the timestamp since it is the first message he
receives.

We say that there is an attack if there is a joined solution of the
\dedsys{} and  the previously mentioned time
constraints together with $x_{t_7}\geq 30$. This last constraint expresses that the 
timestamp received by $B$ is too
large to come from $A$.
Altogether, the time constraint becomes
$x_{t_2}\leq x_{t_1}+30\ \wedge\  x_{t_4}\leq x_{t_3}+30\ \wedge\  x_{t_6}\leq
x_{t_5}+30 \wedge\ x_{t_7}\geq 30.$
Then the substitution corresponding to the attack is
$$\sigma=\set{\subst{y_1}{k_{ab}},\subst{y_2}{k_{ab}},\subst{y_3}{k_{ab}},\subst{y_4}{k_{ab}},
\subst{x_{t_1}}{0},\subst{x_{t_2}}{30},\subst{x_{t_3}}{30},\subst{x_{t_4}}{60},
\subst{x_{t_5}}{60},\subst{x_{t_6}}{90},\subst{x_{t_7}}{90}}.$$
\comment{$$\sigma = \{y_1=y_2=y_3=y_4=k_{ab},\, x_{t_1}\!\!= 0,\, x_{t_2}\!\!=x_{t_3}\!\!=
30,\,  x_{t_4}\!\!=x_{t_5}\!\!= 60,\,
x_{t_6}\!\!=x_{t_7}\!\!=90\}.$$}
\end{example}


\begin{proposition}\label{prop:timed}
There is an attack to a solved \dedsys{} and a time constraint $T$ iff
$T$ has a solution. 
\end{proposition}
\begin{proof}[sketch]
Let $C$ be a solved \dedsys{},  and $T$ a timed constraint. Let
$y_1,\ldots,y_n$ be the variables of sort $\term$ in $C$ and $x_1,\ldots,x_k$ the variables of sort $\Time$
in $C$. Clearly, any substitution $\sigma$ of the form $y_i\sigma=u_i$ where $u_i\in S_i$ for some
$(S_i\Vdash y_i)\in C$ and $x_i\sigma=t_i$ for $t_i$ any constant of sort $\Time$ is a solution of $C$. Let $\sigma'$ be the restriction of $\sigma$ to the timed variables
$x_1,\ldots,x_k$.

 $\sigma$ is an attack for $C$ and $T$ if and only if $\sigma'$ is a solution to $T$. 
Thus there
exists an attack for $C$ and  $T$ if and only if $T$ is satisfiable. 
\end{proof}

\begin{corollary}
Deciding whether a \dedsys{}, together with a time constraint, has a solution
is NP-complete. 
\end{corollary}

\begin{proof}
The NP membership follows from the NP membership of time constraint satisfiability,  
Theorem~\ref{theo:general} and
Proposition~\ref{prop:timed}.

NP-hardness directly follows from the NP-hardness of \dedsys{} solving, considering
an empty timed constraint.
\end{proof}

\section{Conclusions}\label{sec:conclusion}

We have shown how, revisiting the approach of \cite{CS03,RT03TCS}, we
can preserve the set of solutions, instead of only deciding the satisfiability.
We also derived  NP-completeness results for some security properties:
key-cycles, authentication, time constraints. 

Since the constraint-based approach~\cite{CS03,RT03TCS} has
already been implemented in \avispa~\cite{avispa2005}, it is likely that
we can, with only slight efforts, 
adapt this implementation to the case of key cycles and timestamps.

More generally, we would like to take advantage of our result to derive
decision procedures for even more security properties. A typical example
would be the combinations of several properties. Also, we could investigate
non-trace properties such as anonymity or guessing attacks, for which
there are very few decision results (only \cite{BaudetCCS05}, whose procedure
is quite complex). 





Regarding key cycles, our approach is valid for a bounded number of
sessions only. Secrecy is undecidable in
general~\cite{durgin04undecidability} for an unbounded number of
sessions. Such an undecidability result could be easily adapted to the
problem of detecting key cycles.  Secrecy is decidable
for several classes of protocols~\cite{RS03,rta03,blanchetTag03,VermaCade05} 
and an unbounded number of sessions. We plan to investigate how such
fragments could be used to decide key cycles.

\paragraph{Acknowledgments.} We are particularly grateful to Michael Backes, Micha\"el
Rusinowitch, St\'ephanie Delaune, and Bogdan Warinschi for their very helpful  suggestions.

\bibliographystyle{acmtrans}
\bibliography{key_cycles}

\begin{thebibliography}{}

\bibitem[\protect\citeauthoryear{Abadi and Rogaway}{Abadi and
  Rogaway}{2002}]{ARCryptology02}
{\sc Abadi, M.} {\sc and} {\sc Rogaway, P.} 2002.
\newblock Reconciling two views of cryptography (the computational soundness of
  formal encryption).
\newblock {\em Journal of Cryptology\/}~{\em 2}, 103--127.

\bibitem[\protect\citeauthoryear{Ad{\~a}o, Bana, Herzog, and Scedrov}{Ad{\~a}o
  et~al\mbox{.}}{2005}]{AdaoBanaHerzogScedrov-ESORICS05}
{\sc Ad{\~a}o, P.}, {\sc Bana, G.}, {\sc Herzog, J.}, {\sc and} {\sc Scedrov,
  A.} 2005.
\newblock Soundness of formal encryption in the presence of key-cycles.
\newblock In {\em Proc. of the 10th European Symposium on Research in Computer
  Security (ESORICS'05)}. Lecture Notes in Computer Science, vol. 3679.
  Springer Verlag, 374--396.

\bibitem[\protect\citeauthoryear{Amadio and Lugiez}{Amadio and
  Lugiez}{2000}]{AL00}
{\sc Amadio, R.} {\sc and} {\sc Lugiez, D.} 2000.
\newblock On the reachability problem in cryptographic protocols.
\newblock In {\em Proc. of the 11th Int. Conf. on Concurrency Theory
  (CONCUR'00)}. Lecture Notes in Computer Science, vol. 1877. Springer Verlag,
  380--394.

\bibitem[\protect\citeauthoryear{Armando, Basin, Boichut, Chevalier, Compagna,
  Cuellar, Drielsma, H\'eam, Kouchnarenko, Mantovani, M\"odersheim, von Oheimb,
  Rusinowitch, Santiago, Turuani, Vigan\`o, and Vigneron}{Armando
  et~al\mbox{.}}{2005}]{avispa2005}
{\sc Armando, A.}, {\sc Basin, D.}, {\sc Boichut, Y.}, {\sc Chevalier, Y.},
  {\sc Compagna, L.}, {\sc Cuellar, J.}, {\sc Drielsma, P.~H.}, {\sc H\'eam,
  P.}, {\sc Kouchnarenko, O.}, {\sc Mantovani, J.}, {\sc M\"odersheim, S.},
  {\sc von Oheimb, D.}, {\sc Rusinowitch, M.}, {\sc Santiago, J.}, {\sc
  Turuani, M.}, {\sc Vigan\`o, L.}, {\sc and} {\sc Vigneron, L.} 2005.
\newblock The {AVISPA} tool for the automated validation of internet security
  protocols and applications.
\newblock In {\em Proc. of the Computer Aided Verification (CAV'05)}. Lecture
  Notes in Computer Science, vol. 3576. Springer Verlag.

\bibitem[\protect\citeauthoryear{Backes and Pfitzmann}{Backes and
  Pfitzmann}{2004}]{Backes_Pfitzmann_CSFW04_symmetric_encryption}
{\sc Backes, M.} {\sc and} {\sc Pfitzmann, B.} 2004.
\newblock Symmetric encryption in a simulatable {D}olev-{Y}ao style
  cryptographic library.
\newblock In {\em Proc. of the 17th IEEE Computer Security Foundations Workshop
  (CSFW'04)}. IEEE Computer Society Press, 204--218.

\bibitem[\protect\citeauthoryear{Backes, Pfitzmann, and Scedrov}{Backes
  et~al\mbox{.}}{2007}]{backes07key}
{\sc Backes, M.}, {\sc Pfitzmann, B.}, {\sc and} {\sc Scedrov, A.} 2007.
\newblock Key-dependent message security under active attacks --
  {BRSIM/UC}-soundness of symbolic encryption with key cycles.
\newblock In {\em Proc. of the 20th IEEE Computer Security Foundations
  Symposium (CSF'07)}. IEEE Computer Society Press.
\newblock Preprint on IACR ePrint 2005/421.

\bibitem[\protect\citeauthoryear{Baudet}{Baudet}{2005}]{BaudetCCS05}
{\sc Baudet, M.} 2005.
\newblock Deciding security of protocols against off-line guessing attacks.
\newblock In {\em Proc. of the 12th ACM Conf. on Computer and Communication
  Security (CCS'05)}. ACM Press, 16--25.

\bibitem[\protect\citeauthoryear{Bellare and Rogaway}{Bellare and
  Rogaway}{1993}]{BellareRogaway-CRYPTO-1993}
{\sc Bellare, M.} {\sc and} {\sc Rogaway, P.} 1993.
\newblock Entity authentication and key distribution.
\newblock In {\em Proc. of the 13th Annual Int. Conf. on Advances in Cryptology
  (CRYPTO'93)}. Lecture Notes in Computer Science, vol. 773. Springer Verlag,
  232--249.

\bibitem[\protect\citeauthoryear{Blanchet}{Blanchet}{2001}]{Blanchet_CSFW01_ef%
ficient_verifier}
{\sc Blanchet, B.} 2001.
\newblock An efficient cryptographic protocol verifier based on {P}rolog rules.
\newblock In {\em Proc. of the 14th IEEE Computer Security Foundations Workshop
  (CSFW'01)}. IEEE Computer Society Press, 82--96.

\bibitem[\protect\citeauthoryear{Blanchet and Podelski}{Blanchet and
  Podelski}{2003}]{blanchetTag03}
{\sc Blanchet, B.} {\sc and} {\sc Podelski, A.} 2003.
\newblock Verification of cryptographic protocols: Tagging enforces
  termination.
\newblock In {\em Foundations of Software Science and Computation Structures
  (FoSSaCS'03)}, {A.~Gordon}, Ed. Lecture Notes in Computer Science, vol. 2620.
  Springer Verlag, 136--152.

\bibitem[\protect\citeauthoryear{Bozga, Ene, and Lakhnech}{Bozga
  et~al\mbox{.}}{2004}]{BEL04concur}
{\sc Bozga, L.}, {\sc Ene, C.}, {\sc and} {\sc Lakhnech, Y.} 2004.
\newblock A symbolic decision procedure for cryptographic protocols with time
  stamps.
\newblock In {\em Proc. of the 15th Int. Conf. on Concurrency Theory
  (CONCUR'04)}. Lecture Notes in Computer Science, vol. 3170. Springer Verlag,
  177--192.

\bibitem[\protect\citeauthoryear{Bursuc, Comon{-}Lundh, and Delaune}{Bursuc
  et~al\mbox{.}}{2007}]{BCD-stacs2007}
{\sc Bursuc, S.}, {\sc Comon{-}Lundh, H.}, {\sc and} {\sc Delaune, S.} 2007.
\newblock Associative-commutative deducibility constraints.
\newblock In {\em Proc. of the 24th Annual Symposium on Theoretical Aspects of
  Computer Science (STACS'07)}. Lecture Notes in Computer Science, vol. 4393.
  Springer Verlag, 634--645.

\bibitem[\protect\citeauthoryear{Clark and Jacob}{Clark and Jacob}{1997}]{CJ97}
{\sc Clark, J.} {\sc and} {\sc Jacob, J.} 1997.
\newblock A survey of authentication protocol literature.
\newblock Available at
  \url{http://www.cs.york.ac.uk/~jac/papers/drareviewps.ps}.

\bibitem[\protect\citeauthoryear{Colmerauer}{Colmerauer}{1984}]{colmerauer84}
{\sc Colmerauer, A.} 1984.
\newblock Equations and inequations on finite and infinite trees.
\newblock In {\em Proc. of the Int. Conf. on Fifth Generation Computer Systems
  (FGCS'84)}. 85--99.

\bibitem[\protect\citeauthoryear{Comon-Lundh and Cortier}{Comon-Lundh and
  Cortier}{2003}]{rta03}
{\sc Comon-Lundh, H.} {\sc and} {\sc Cortier, V.} 2003.
\newblock New decidability results for fragments of first-order logic and
  application to cryptographic protocols.
\newblock In {\em Proc. of the 14th Int. Conf. on Rewriting Techniques and
  Applications (RTA'03)}. Lecture Notes in Computer Science, vol. 2706.
  Springer Verlag, 148--164.

\bibitem[\protect\citeauthoryear{Comon-Lundh and Shmatikov}{Comon-Lundh and
  Shmatikov}{2003}]{CS03}
{\sc Comon-Lundh, H.} {\sc and} {\sc Shmatikov, V.} 2003.
\newblock Intruder deductions, constraint solving and insecurity decision in
  presence of exclusive or.
\newblock In {\em Proc. of the 18th Annual IEEE Symposium on Logic in Computer
  Science (LICS'03)}. IEEE Computer Society Press, 271--280.

\bibitem[\protect\citeauthoryear{Corin}{Corin}{2006}]{Corin-thesis}
{\sc Corin, R.} 2006.
\newblock Analysis models for security protocols.
\newblock Ph.D. thesis, University of Twente, The Netherlands.

\bibitem[\protect\citeauthoryear{Corin and Etalle}{Corin and
  Etalle}{2002}]{CorinE02}
{\sc Corin, R.} {\sc and} {\sc Etalle, S.} 2002.
\newblock An improved constraint-based system for the verification of security
  protocols.
\newblock In {\em Proc. of the 9th Int. Symposium on Static Analysis (SAS'02)}.
  Lecture Notes in Computer Science, vol. 2477. Springer Verlag, 326--341.

\bibitem[\protect\citeauthoryear{Corin, Saptawijaya, and Etalle}{Corin
  et~al\mbox{.}}{2005}]{corin-psltl}
{\sc Corin, R.~J.}, {\sc Saptawijaya, A.}, {\sc and} {\sc Etalle, S.} 2005.
\newblock {PS-LTL} for constraint-based security protocol analysis.
\newblock In {\em Proc. of the 21st Int. Conf. on (ICLP'05)}. Lecture Notes in
  Computer Science, vol. 3668. Springer Verlag, 439--440.

\bibitem[\protect\citeauthoryear{Cortier, Delaitre, and Delaune}{Cortier
  et~al\mbox{.}}{2007}]{cortier07fsttcs}
{\sc Cortier, V.}, {\sc Delaitre, J.}, {\sc and} {\sc Delaune, S.} 2007.
\newblock Safely composing security protocols.
\newblock In {\em Proc. of the 27th Int. Conf. on Foundations of Software
  Technology and Theoretical Computer Science (FSTTCS'07)}. Lecture Notes in
  Computer Science, vol. 4855. Springer Verlag, 352--363.

\bibitem[\protect\citeauthoryear{Cortier, Kremer, K{\"u}sters, and
  Warinschi}{Cortier et~al\mbox{.}}{2006}]{CKKW-fsttcs2006}
{\sc Cortier, V.}, {\sc Kremer, S.}, {\sc K{\"u}sters, R.}, {\sc and} {\sc
  Warinschi, B.} 2006.
\newblock Computationally sound symbolic secrecy in the presence of hash
  functions.
\newblock In {\em Proc. of the 26th Int. Conf. on Foundations of Software
  Technology and Theoretical Computer Science (FSTTCS'06)}. Lecture Notes in
  Computer Science, vol. 4337. Springer Verlag, 176--187.

\bibitem[\protect\citeauthoryear{Cortier and Z\u{a}linescu}{Cortier and
  Z\u{a}linescu}{2006}]{CortierLPAR06}
{\sc Cortier, V.} {\sc and} {\sc Z\u{a}linescu, E.} 2006.
\newblock Deciding key cycles for security protocols.
\newblock In {\em Proc. of the 13th Int. Conf. on Logic for Programming,
  Artificial Intelligence, and Reasoning (LPAR'06)}. Lecture Notes in
  Artificial Intelligence, vol. 4246. Springer Verlag, 317--331.

\bibitem[\protect\citeauthoryear{Cremers}{Cremers}{2008}]{cremers08cav}
{\sc Cremers, C.} 2008.
\newblock The {S}cyther {T}ool: Verification, falsification, and analysis of
  security protocols.
\newblock In {\em Proc. of the 20th Int. Conf. Computer Aided Verification
  (CAV'08)}. Lecture Notes in Computer Science, vol. 5123. Springer Verlag,
  414--418.

\bibitem[\protect\citeauthoryear{Durgin, Lincoln, and Mitchell}{Durgin
  et~al\mbox{.}}{2004}]{durgin04undecidability}
{\sc Durgin, N.}, {\sc Lincoln, P.}, {\sc and} {\sc Mitchell, J.} 2004.
\newblock Multiset rewriting and the complexity of bounded security protocols.
\newblock {\em Journal of Computer Security\/}~{\em 12,\/}~2, 247--311.

\bibitem[\protect\citeauthoryear{Durgin, Lincoln, Mitchell, and Scedrov}{Durgin
  et~al\mbox{.}}{1999}]{durgin99undecidability}
{\sc Durgin, N.}, {\sc Lincoln, P.}, {\sc Mitchell, J.}, {\sc and} {\sc
  Scedrov, A.} 1999.
\newblock Undecidability of bounded security protocols.
\newblock In {\em Proc. of the Workshop on Formal Methods and Security
  Protocols}.

\bibitem[\protect\citeauthoryear{Goldwasser and Micali}{Goldwasser and
  Micali}{1984}]{Goldwasser_Micali_JCSS84_probabilistic_encryption}
{\sc Goldwasser, S.} {\sc and} {\sc Micali, S.} 1984.
\newblock Probabilistic encryption.
\newblock {\em Journal of Computer and System Sciences\/}~{\em 28}, 270--299.

\bibitem[\protect\citeauthoryear{Hofheinz and Unruh}{Hofheinz and
  Unruh}{2008}]{hofheinz08towards}
{\sc Hofheinz, D.} {\sc and} {\sc Unruh, D.} 2008.
\newblock Towards key-dependent message security in the standard model.
\newblock In {\em EUROCRYPT 2008}. Lecture Notes in Computer Science, vol.
  4965. Springer Verlag, 108--126.
\newblock Preprint on IACR ePrint 2007/333.

\bibitem[\protect\citeauthoryear{Janvier}{Janvier}{2006}]{Janvier-these}
{\sc Janvier, R.} 2006.
\newblock Lien entre mod\`eles symboliques et computationnels pour le
  protocoles cryptographiques utilisant des hachage.
\newblock Ph.D. thesis, Universit\'e Joseph Fourier, Grenoble.

\bibitem[\protect\citeauthoryear{Janvier, Lakhnech, and Mazare}{Janvier
  et~al\mbox{.}}{2005}]{cryptoeprint:2005:020}
{\sc Janvier, R.}, {\sc Lakhnech, Y.}, {\sc and} {\sc Mazare, L.} 2005.
\newblock {(De)Compositions of Cryptographic Schemes and their Applications to
  Protocols}.
\newblock Cryptology ePrint Archive, Report 2005/020.

\bibitem[\protect\citeauthoryear{Laud}{Laud}{2002}]{Laud-NORDSEC02}
{\sc Laud, P.} 2002.
\newblock Encryption cycles and two views of cryptography.
\newblock In {\em Proc. of the Nordic Workshop on Secure IT Systems
  (NORDSEC'02)}.

\bibitem[\protect\citeauthoryear{Lowe}{Lowe}{1996}]{lowe96breaking}
{\sc Lowe, G.} 1996.
\newblock Breaking and fixing the {N}eedham-{S}chroeder public-key protocol
  using {FDR}.
\newblock In {\em Proc. of the 2nd Int. Workshop on Tools and Algorithms for
  the Construction and Analysis of Systems (TACAS'96)}. Lecture Notes in
  Computer Science, vol. 1055. Springer Verlag, 147--166.

\bibitem[\protect\citeauthoryear{Lowe}{Lowe}{1998}]{lowe98towards}
{\sc Lowe, G.} 1998.
\newblock Towards a completeness result for model checking of security
  protocols.
\newblock In {\em Proc. of the 11th IEEE Computer Security Foundations Workshop
  (CSFW'98)}. IEEE Computer Society Press.

\bibitem[\protect\citeauthoryear{Micciancio and Warinschi}{Micciancio and
  Warinschi}{2004a}]{MW04}
{\sc Micciancio, D.} {\sc and} {\sc Warinschi, B.} 2004a.
\newblock Completeness theorems for the {Abadi-Rogaway} logic of encrypted
  expressions.
\newblock {\em Journal of Computer Security\/}~{\em 12,\/}~1, 99--129.
\newblock Preliminary version in WITS'02.

\bibitem[\protect\citeauthoryear{Micciancio and Warinschi}{Micciancio and
  Warinschi}{2004b}]{Micciancio_Warinschi_TCC04_soundness_of_formal_encryption}
{\sc Micciancio, D.} {\sc and} {\sc Warinschi, B.} 2004b.
\newblock Soundness of formal encryption in the presence of active adversaries.
\newblock In {\em Proc. of the 1st Theory of Cryptography Conference (TCC'04)}.
  Lecture Notes in Computer Science, vol. 2951. Springer Verlag, 133--151.

\bibitem[\protect\citeauthoryear{Millen and Shmatikov}{Millen and
  Shmatikov}{2001}]{MS01}
{\sc Millen, J.} {\sc and} {\sc Shmatikov, V.} 2001.
\newblock Constraint solving for bounded-process cryptographic protocol
  analysis.
\newblock In {\em Proc. of the 8th ACM Conf. on Computer and Communication
  Security (CCS'01)}. ACM Press, 166--175.

\bibitem[\protect\citeauthoryear{Needham and Schroeder}{Needham and
  Schroeder}{1978}]{NS78}
{\sc Needham, R.~M.} {\sc and} {\sc Schroeder, M.~D.} 1978.
\newblock Using encryption for authentication in large networks of computers.
\newblock {\em Communications of the ACM\/}~{\em 21,\/}~12, 993--999.

\bibitem[\protect\citeauthoryear{Ramanujam and Suresh}{Ramanujam and
  Suresh}{2003}]{RS03}
{\sc Ramanujam, R.} {\sc and} {\sc Suresh, S.~P.} 2003.
\newblock Tagging makes secrecy decidable for unbounded nonces as well.
\newblock In {\em Proc. of the 23rd Conf. on Foundations of Software Technology
  and Theoretical Computer Science (FSTTCS'03)}. Lecture Notes in Computer
  Science, vol. 2914. Springer Verlag, 363--374.

\bibitem[\protect\citeauthoryear{Ramanujam and Suresh}{Ramanujam and
  Suresh}{2005}]{ramanujam05jcs}
{\sc Ramanujam, R.} {\sc and} {\sc Suresh, S.~P.} 2005.
\newblock Decidability of context-explicit security protocols.
\newblock {\em Journal of Computer Security\/}~{\em 13,\/}~1, 135--165.

\bibitem[\protect\citeauthoryear{Rusinowitch and Turuani}{Rusinowitch and
  Turuani}{2001}]{RT01}
{\sc Rusinowitch, M.} {\sc and} {\sc Turuani, M.} 2001.
\newblock Protocol insecurity with finite number of sessions is {NP}-complete.
\newblock In {\em Proc. of the 14th IEEE Computer Security Foundations Workshop
  (CSFW'01)}. IEEE Computer Society Press, 174--190.

\bibitem[\protect\citeauthoryear{Rusinowitch and Turuani}{Rusinowitch and
  Turuani}{2003}]{RT03TCS}
{\sc Rusinowitch, M.} {\sc and} {\sc Turuani, M.} 2003.
\newblock Protocol insecurity with finite number of sessions and composed keys
  is {NP}-complete.
\newblock {\em Theoretical Computer Science\/}~{\em 299}, 451--475.

\bibitem[\protect\citeauthoryear{Syverson and Meadows}{Syverson and
  Meadows}{1996}]{SyversonM96}
{\sc Syverson, P.} {\sc and} {\sc Meadows, C.} 1996.
\newblock A formal language for cryptographic protocol requirements.
\newblock {\em Designes, Codes and Cryptography\/}~{\em 7,\/}~1-2, 27--59.

\bibitem[\protect\citeauthoryear{Verma, Seidl, and Schwentick}{Verma
  et~al\mbox{.}}{2005}]{VermaCade05}
{\sc Verma, K.~N.}, {\sc Seidl, H.}, {\sc and} {\sc Schwentick, T.} 2005.
\newblock On the complexity of equational {H}orn clauses.
\newblock In {\em Proc. of the 22th Int. Conf. on Automated Deduction
  (CADE'05)}. Lecture Notes in Computer Science. Springer Verlag, 337--352.

\end{thebibliography}

\begin{received}
Received August 2007;
accepted April 2008
\end{received}

\end{document}